\documentclass[journal=jpcl,manuscript=extarticle,layout=twocolumn]{achemso}
\setkeys{acs}{maxauthors=0}

\let\titlefont\undefined
\makeatletter
\let\l@addto@macro\relax
\makeatother
\usepackage[fontsize=9pt]{scrextend}

\usepackage{graphicx}% Include figure files
\usepackage{bm}% bold math
\usepackage{xcolor}
\usepackage{siunitx}
\usepackage{physics}

\usepackage{soul} % provides st for strikethrough
\newcommand{\rev}[1]{\textcolor{black}{#1}}

\usepackage{amssymb}

\renewcommand{\arraystretch}{1.5}
\usepackage[symbol]{footmisc}

\renewcommand{\thefootnote}{\fnsymbol{footnote}}

\newcommand{\activestate}{n}

\usepackage{braket}
\usepackage{dcolumn} % D
\newcolumntype{d}[1]{D{.}{.}{#1}} % taXble environment which aligns decimal places

\title{Recovering Marcus Theory Rates and Beyond without the Need for Decoherence Corrections: The Mapping Approach to Surface Hopping}

\author{Joseph E.\ Lawrence}
\affiliation{Department of Chemistry and Applied Biosciences, ETH Zurich, 8093 Zurich, Switzerland}
\email{joseph.lawrence@phys.chem.ethz.ch}
\author{Jonathan R.\ Mannouch}
\affiliation{Hamburg Center for Ultrafast Imaging, Universit\"at Hamburg and the Max Planck Institute for the Structure and Dynamics of Matter, Luruper Chaussee 149, 22761 Hamburg, Germany}
\email{jonathan.mannouch@mpsd.mpg.de}
\author{Jeremy~O.\ Richardson}
\affiliation{Department of Chemistry and Applied Biosciences, ETH Zurich, 8093 Zurich, Switzerland}
\email{jeremy.richardson@phys.chem.ethz.ch}

\date{\today}% It is always \today, today,
\makeatletter
\newcommand*{\addFileDependency}[1]{
  \typeout{(#1)}
  \@addtofilelist{#1}
  \IfFileExists{#1}{}{\typeout{No file #1.}}
}
\makeatother

\begin{document}

\maketitle

\begin{center}
\includegraphics{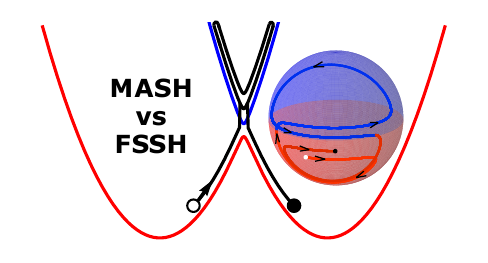}
\end{center}
\textbf{ABSTRACT:} It is well known that fewest-switches surface hopping (FSSH) fails to correctly capture the quadratic scaling of rate constants with diabatic coupling in the weak-coupling limit, as expected from Fermi's golden rule and Marcus theory. To address this deficiency, the most widely used approach is to introduce a `decoherence correction', 
which removes the inconsistency between the wavefunction coefficients and the active state. %, such as is done in the A-FSSH approach.
Here we investigate the behavior of a new nonadiabatic trajectory method, called the mapping approach to surface hopping (MASH), on systems that exhibit incoherent rate behavior.
Unlike FSSH, MASH hops between active surfaces deterministically, and can never have an inconsistency between the wavefunction coefficients and the active state.
We show that MASH is not only able to describe rates for intermediate and strong diabatic coupling, but can also accurately reproduce the results of Marcus theory in the golden-rule limit, without the need for a decoherence correction.
MASH is therefore a significant improvement over FSSH in the simulation of nonadiabatic reactions.

\textbf{Introduction:} 
Under the Born--Oppenheimer approximation, one assumes that electronic motion is fast compared to nuclear motion and is therefore adiabatically separated. The resulting picture of nuclei moving on a single adiabatic potential energy surface
 forms the basis of our modern understanding of molecular structure and dynamics.
Despite its great success, there are many important molecular processes for which the Born--Oppenheimer approximation is not valid. Most obviously this can occur in processes, such as photo-excitation, where the electronic degrees of freedom are driven far from equilibrium.\cite{Worth2004BeyondBO,Curchod2018review,Barbatti2011FSSHReview,Subotnik2016review,Gomez2020IntroChapterNonAdDyn,Mai2020FSSHChapter} However, nonadiabatic dynamics can also occur closer to equilibrium in processes which involve significant redistribution of electron density, such as in electron transfer.\cite{Marcus1956ET,Marcus1985review,Bader1990golden,HammesSchiffer2015PCET,Toldo2023FSSH_ET_persp} The importance of both light--matter interaction as well as electron-transfer processes to physics, chemistry and biology as well as modern technology makes the development of practical simulation methods for nonadiabatic dynamics of utmost importance.\cite{HammesSchiffer2010PCET,Akimov2014ChargeTransferSingletFission,Blumberger2015ET}

Unfortunately, finding an exact solution of the full coupled electron--nuclear Schr\"odinger equation is impractical for most systems of interest, and hence approximations need to be made.\cite{Stock2005nonadiabatic,Kelly2013SH,Martens2016CSH,Min2015nonadiabatic,Ha2018XFSH} Fortunately, however, the relatively high mass of atomic nuclei means that it is often a reasonable approximation to treat them as classical particles with well-defined positions and momenta. In 1990 Tully proposed what has become the most widely used of such `mixed quantum--classical' methods for simulating nonadiabatic processes, known as fewest-switches surface hopping (FSSH).\cite{Tully1990hopping} 
Within FSSH, the nuclei predominantly move under the force of a single adiabatic potential energy surface, with occasional  stochastic hops between the surfaces. The probabilities for these hopping events are determined based on the evolution of the electronic wavefunction under the time-dependent Hamiltonian generated by the nuclear trajectory.

Fewest-switches surface hopping has been successfully applied to study a wide range of nonadiabatic processes.\cite{Barbatti2011FSSHReview,Subotnik2016review,Gomez2020IntroChapterNonAdDyn,Mai2020FSSHChapter} However, it has long been appreciated that
there are problems %, due to for instance frustrated hops, 
that lead to a breakdown in the assumptions behind the FSSH algorithm.\cite{Schwartz1994ET,Bittner1995Decoherence,Schwartz1996Decoherence,Fang1999FSSH_inconsistency} 
The result is a deviation between the number of trajectories on each surface and the wavefunction coefficients, which can therefore be referred to as an inconsistency error. At a more fundamental level, the error can be attributed to a failure to describe the decoherence of the electronic wavefunction that results from the splitting of a wavepacket after passing through a coupling region.\cite{Schwartz1994ET,Bittner1995Decoherence,Schwartz1996Decoherence} This observation has led to the introduction of many different ad-hoc decoherence corrections, aimed at fixing the inconsistency (overcoherence) error of FSSH.\cite{Fang1999FSSH_inconsistency,Schwartz1996Decoherence,Wong2002Decoherence,Wong2002Decoherence2,Jasper2005Decoherence,Subotnik2011AFSSH,Jain2016AFSSH}

Due to their ad-hoc nature, decoherence corrections are not guaranteed to consistently improve the results of a calculation.\cite{Plasser2019DecoherenceMomentumRescaling} However, one area where they have been shown to be essential is for processes, such as electron transfer, which involve slow population transfer in strongly nonadiabatic systems (weak diabatic coupling, $\Delta$). A series of papers from Subotnik and coworkers has demonstrated that the standard FSSH algorithm fails to properly describe the $\Delta^2$ scaling of the rate predicted by Fermi's golden-rule and the famous Marcus theory of electron transfer.\cite{Landry2011hopping,Landry2012hopping,Jain2015hopping1,Jain2015hopping2,Falk2014FSSHFriction} This was explained in terms of repeated crossings of the nonadiabatic coupling region, leading to a build up of the inconsistency error.\cite{Landry2011hopping}

Recently, an alternative to FSSH has been derived known as the mapping approach to surface hopping (MASH).\cite{Mannouch2023MASH} MASH was designed to offer the best of both worlds between surface hopping and mapping approaches, such as the Meyer--Miller--Stock--Thoss mapping\cite{Meyer1979nonadiabatic,Stock1997mapping} and spin mapping.\cite{spinmap,multispin} Unlike FSSH, which was proposed heuristically, MASH can be rigorously derived from the quantum--classical Liouville equation (QCLE).\cite{Kapral1999MQCD,Shi2004QCLE,Bonella2010QCLE,Kelly2012mapping,Subotnik2013QCLE,Kapral2016FSSH} Tests against exact results for the Tully models, a series of spin-boson models, as well as 3-mode and 24-mode vibronic models of
pyrazine %\cite{Raab1999Pyrazine,Chen2006Pyrazine,Baiardi2019Pyrazine}
have shown that the results of MASH are generally as good or better than FSSH for an equivalent computational cost.\cite{Mannouch2023MASH} %Additionally, the rigorous connection of MASH to the QCLE means that by applying quantum jumps it is possible to systematically improve the results towards those of the QCLE. One can also rigorously derive decoherence corrections. Interestingly however it was found that these are needed as often as one might expect.... 
Perhaps most interesting are the results for the spin-boson model where the system crosses the coupling region many times during the dynamics. One might have expected that decoherence corrections were necessary to improve upon the FSSH results. However, MASH shows a significant improvement even without the addition of decoherence corrections. This raises the question, how well will MASH perform in systems exhibiting slow population transfer with weak diabatic coupling where the errors of FSSH are known to be particularly pronounced?\cite{Landry2011hopping}

In the following we will attempt to answer this question. In doing so, we will explore  the difference between MASH and FSSH in terms of the language of decoherence, revisiting the reasons for the breakdown of FSSH in systems with weak diabatic couplings, and showing how MASH improves upon these issues. We will begin by giving an overview of the two methods, highlighting the key similarities and differences between the FSSH and MASH algorithms. We will then describe how to simulate nonadiabatic rates using these approaches, before a detailed discussion of how each of the methods performs for a range of different physically relevant parameter regimes. 

\textbf{Methods:} Here we give a brief description of the two methods in the case of a two-level system. 
Both FSSH and MASH treat the nuclear motion classically, with the nuclear positions and momenta represented by the classical variables $\bm{q}(t)$ and $\bm{p}(t)$ respectively. %In each time step
Between hopping events, the nuclei evolve under a force which is given by the derivative of the adiabatic potential corresponding to the `active surface'
\begin{equation}
    \bm{F}= - \frac{\partial V_{\activestate}}{\partial \bm{q}},
\end{equation}
where $\activestate$ is the active-state variable, and we label the upper adiabat $+$ and the lower adiabat $-$. Electronic wavefunction coefficients, $c_\pm(t)$, are then propagated according to the time-dependent Schr\"odinger equation under the Hamiltonian generated by the nuclear trajectory.
In both theories, these coefficients are used to determine when to hop, but 
 are not used to calculate adiabatic population observables, which are instead obtained directly from the fraction of trajectories on a given active surface.\cite{Subotnik2016review} 
An intuitive picture of the electronic dynamics can be obtained using the coordinates of the Bloch sphere
\begin{subequations}
\begin{alignat}{3}
     S_x &= c_+c^*_-+c_+^*c_- \\
     S_y &= i[c_+c^*_--c_+^*c_-] \\
     S_z &= |c_+|^2-|c_-|^2.
\end{alignat}
\end{subequations}
This highlights the equivalence of the electronic dynamics to the rotation of a classical spin around a magnetic field
\begin{equation}
    \hbar \dot{\bm{S}} = \begin{pmatrix} 0 \\ \sum_\mu\frac{2\hbar}{m_\mu}d_\mu(\bm{q})p_\mu \\ V_+(\bm{q})-V_-(\bm{q}) \end{pmatrix} \times \begin{pmatrix} S_x \\ S_y \\ S_z \end{pmatrix},
\end{equation}
where $V_\pm$ are the potentials corresponding to adiabatic states $\phi_{\pm}$, and $d_\mu=\left\langle{\phi_+}\middle|{\frac{\partial \phi_-}{\partial q_\mu}}\right\rangle$ is the nonadiabatic coupling vector.

What differs between FSSH and MASH is how the hops between surfaces are determined.
Within FSSH, the probability of hopping from one surface to the other in a time-step $\delta t$ is given by 
\begin{subequations}
\begin{equation}
    P_{-\to+} =  \frac{\frac{\partial}{\partial t}|c_+(t)|^2 }{|c_-(t)|^2} \delta t = \frac{\dot{S}_z(t)}{1-S_z(t)} \delta t   
\end{equation}
\begin{equation}
        P_{+\to-} =  \frac{\frac{\partial}{\partial t}|c_-(t)|^2 }{|c_+(t)|^2} \delta t = \frac{-\dot{S}_z(t)}{1+S_z(t)} \delta t ,
\end{equation}
\end{subequations}
where negative probabilities indicate no hop. In contrast to this, the active surface in MASH is obtained deterministically by the simple condition
\begin{equation}
   \activestate(t) = \mathrm{sign}\left(|c_{+}(t)|^2 - |c_{-}(t)|^2\right)=\mathrm{sign}(S_z(t)),
\end{equation}
i.e.~the active state is the one with the larger probability, $|c_{\pm}(t)|^2$.  The fact that MASH is deterministic might seem surprising, particularly given that it is the stochastic nature of FSSH that allows it to describe wavepacket splitting. However, as in other mapping-based methods,\cite{spinmap} the stochastic nature of surface hopping is replaced in MASH by sampling over initial wavefunction coefficients, as we shall explain below. 
To complete the specification of the dynamics, we need to define
what happens to the momentum at a hopping (or attempted hopping) event.
While there has been some debate in the literature as to how this should be done in FSSH,\cite{Muller1997FSSH,Jasper2001Reversal} the derivation of MASH from the QCLE leads to a unique prescription for how to deal with momentum rescaling and so-called frustrated hops (where the trajectory has insufficient energy to hop). %This happens to be
The result is equivalent to what was originally argued for by Tully\cite{HammesSchiffer1994FSSH} (along with many others\cite{Pechukas1969scattering2,Herman1984MomentumReversal}): the momenta are rescaled along the direction of the nonadiabatic coupling and are reflected in all cases that they do not have sufficient energy to hop.  

This suffices to describe the dynamical evolution of MASH and FSSH, however there is one additional important difference: how the simulation is initialized. For ease of comparison between FSSH and MASH, we will focus here on the calculation of correlation functions that involve only adiabatic populations and nuclear configurations (although we note that the MASH derivation leads to a rigorous prescription for the calculation of correlation functions involving electronic coherences). For a system starting in a specific adiabatic state, both FSSH and MASH are initialized with the corresponding active state, $\activestate(0)$.  In FSSH the wavefunction coefficients are initialized as the corresponding pure state, e.g.\ if the initial state is $\activestate=+$ then $c_+(0)=1$ and $c_-(0)=0$ and the initial $\bm{S}$ vector points to the north pole of the Bloch sphere.
In contrast the wavefunction coefficients in MASH are sampled such that the initial $\bm{S}$ is distributed over the entire hemispherical surface of the Bloch sphere corresponding to the initial state, with a probability density proportional to $|S_z|$. It is this sampling that effectively replaces the stochastic nature of the hops in FSSH.

\textbf{Rate Calculations:} Full details of the calculation of rate constants with MASH and FSSH are discussed in the supporting information. Here we give an overview of the most important aspects of reaction rate theory, focusing on the advantages of MASH over FSSH in two key areas: efficiency and accuracy.

Typically, the accurate determination of rate constants from a direct simulation of the population dynamics
is not possible, as the barrier crossing is a rare event and prohibitively long trajectories would be required to observe a statistically significant number of reactions.
The standard approach used to overcome this problem is the flux-correlation formalism.\cite{ChandlerGreen} This avoids the rare-event problem by reformulating the rate in terms of a correction to transition state theory: the transmission coefficient. Importantly, the calculation of the transmission coefficient only involves running a short simulation up to the `plateau' time, $t_{\rm pl}$, which is much shorter than the timescale of the reaction, $t_{\rm pl}\ll\tau_{\rm rxn}$, but long enough that the initial transient behavior has subsided and the population decay is exponential.\cite{ChandlerGreen} 

 %computing the transition state theory rate and then a dynamical correction (the transmission coefficient) by running dynamics initiated at the dividing surface separating reactants and products. Importantly this dynamics only needs to be run for a relatively short time, up to the point at which the transmission coefficient plateaus. 
Unfortunately, the FSSH dynamics do not obey time-translation symmetry, and hence the flux-correlation formalism does not rigorously give the same result as calculating the rate from direct population dynamics. 
A number of approaches to overcome this issue have been suggested, such as using initial wavefunction amplitudes in the flux-correlation function generated from approximate backwards-propagation schemes,\cite{HammesSchiffer1995FSSH_TST,Jain2015hopping2} as well as the use of dynamical enhanced sampling in the form of forward-flux sampling.\cite{Reiner2023ForwardFlux_SH}  
Here, to avoid making further approximations, we simply calculate the FSSH rate from direct population dynamics, which is achievable due to the low computational cost of the model employed.
The calculation of reaction rates with MASH presents a significant advantage in this regard: the dynamics of MASH \emph{do} rigorously obey time-translation symmetry. This means that all of the usual machinery of the flux-correlation formalism (such as the Bennett--Chandler method\cite{Bennett1977TST,Chandler1978TST,MolSim}) can be used to improve the efficiency of rate calculations in a way that is rigorously equivalent to the rate that would be obtained (less efficiently) with a direct simulation of the population dynamics.

The second difficulty associated with the calculation of reaction rates with FSSH is the overcoherence error.\cite{Landry2011hopping,Landry2012hopping,Jain2015hopping1,Jain2015hopping2,Falk2014FSSHFriction} This error is known to occur in problems where the system passes through regions of strong nonadiabatic coupling (equivalent to weak diabatic coupling) multiple times, resulting in an active state that is inconsistent with the wavefunction coefficients.  Importantly, the dynamics of MASH can never become inconsistent in the way they do in FSSH, as the active state is determined explicitly from the wavefunction coefficients. This means that one may expect the overcoherence error to be less significant in MASH than it is in FSSH. %, and this is what we shall demonstrate below. 
In order to assess this, we consider the MASH and FSSH dynamics in two different regimes.
Firstly, we focus on how the error affects dynamics near the plateau time, $t\sim t_{\rm pl}$. This is done by calculating rates from the slope of the product population $\langle P_p(t)\rangle$ after the initial transient behavior has subsided for a system initialized in the reactant well in a classical thermal distribution. %\footnote{For the precise definition of the initial conditions see the supporting information.}
Secondly, we consider the dynamics over the timescale of the reaction, $t\sim\tau_{\rm rxn}$, by simulating the full population decay.

\textbf{Model:} In order to compare numerically the accuracy of MASH and FSSH for the calculation of nonadiabatic rates, we consider the prototypical model for electron transfer: the spin-boson model.\cite{Nitzan} For ease of interpretation, we will consider the Brownian-oscillator form of the spin-boson model,\cite{Leggett1984spinboson,Garg1985spinboson,Thoss2001hybrid,Lawrence2019ET} which consists of a harmonic (mass-weighted) solvent polarisation coordinate, $Q$, and an Ohmic bath describing the effect of friction along $Q$, with spectral density $J(\omega)=\gamma\omega$. The diabatic potentials along the solvent polarisation coordinate are then the famous Marcus parabolas\cite{Marcus1985review}
\begin{subequations}
\begin{alignat}{1}
     U_0(Q) &= \frac{1}{2}\Omega^2\left(Q+\sqrt{\frac{\Lambda}{2\Omega^2}}\right)^2 + \frac{\varepsilon}{2}\\ 
     U_1(Q) &= \frac{1}{2}\Omega^2\left(Q-\sqrt{\frac{\Lambda}{2\Omega^2}}\right)^2 - \frac{\varepsilon}{2}, 
\end{alignat}
\end{subequations}
where $\Lambda$ is the Marcus reorganisation energy, $\varepsilon$ is the reaction driving force, and $\Omega$ is the characteristic frequency of the parabola. The two diabatic states are coupled by a constant diabatic coupling $\Delta$, and the resulting adiabatic potentials along the solvent polarisation coordinate are given by
\begin{equation}
    V_{\pm}(Q) = \frac{U_0(Q)+U_1(Q)}{2}\pm\sqrt{\left(\frac{U_0(Q)-U_1(Q)}{2}\right)^2+\Delta^2}.
\end{equation}

In both FSSH and MASH simulations, the initial positions and momenta are sampled from the classical Boltzmann distributions (not Wigner functions), and the initial active state is chosen with the associated Boltzmann weighting.
As the nuclei are classical, the coupling of the solvent polarisation coordinate, $Q$, to its environment can be implemented efficiently using a Langevin equation with the friction coefficient $\gamma$. Note this is formally equivalent to explicitly simulating the full  multidimensional bath.\cite{Leggett1984spinboson,Garg1985spinboson,Thoss2001hybrid,Lawrence2019ET}

In the limit of weak diabatic coupling ($\Delta\to0$), Marcus theory
predicts that the rate to go from one well to the other is given by\cite{Marcus1985review}
\begin{equation}
  k_{\rm MT} = \frac{\Delta^2}{\hbar} \sqrt{\frac{\pi\beta}{\Lambda}} \exp\left[-\beta \frac{(\Lambda-\varepsilon)^2}{4\Lambda}\right],
\end{equation}
where $\beta=1/k_{\rm B}T$ is the inverse temperature.
Importantly Marcus theory is exact for this model in the weak-coupling limit under the assumption that the nuclear motion can be treated classically, i.e.\ in the absence of nuclear quantum effects such as zero-point energy and tunneling.
This makes Marcus theory a very useful benchmark for assessing the accuracy of FSSH and MASH, which also assume the nuclear motion can be treated classically. In order to assess their behavior for intermediate values of $\Delta$, where Marcus theory is not applicable, numerically exact quantum-mechanical rates were calculated using the hierarchical equations of motion (HEOM).\cite{Tanimura1989HEOM,Tanimura2020HEOM}  All HEOM calculations were performed using the HEOM-Lab code\cite{Fay2022LowTempHEOM,heomlab} following the method described in Refs.~\citenum{Lawrence2019ET} and \citenum{Shi2009Zusman}. %Parameters were chosen to be as close to the classical nuclear limit as practically achievable with reasonable computational resources. 

For both MASH and FSSH, the long-time behavior of $\langle P_p(t)\rangle$ is independent of the precise definition of reactants and products.%The long-time population dynamics of MASH and FSSH are independent of the precise definition of reactants and products.
\footnote{This is because the parts of the thermal distribution corresponding to reactants and products are two regions of high probability density that are well separated in phase space. Hence, any definition that respects this separation will give the same long-time dynamics. This is true independent of whether or not MASH and FSSH obey detailed balance globally provided they do so in the reactant and product wells, which is trivially the case when the system is electronically adiabatic in these regions. %We note that while MASH does not rigorously obey detailed balance globally, it can be shown that it does thermalize correctly in the long-time limit.\cite{thermalizationarxiv}
}
However, the definition of reactants and products will affect its short-time behavior. The optimum choice for the calculation of rates is the one for which the dynamics of a system initialized in the reactants most quickly settles into exponential decay.
 Normally this is a purely practical matter, however, choosing a (near) optimal definition has an additional importance in the present study: it allows us to separate the short- and long-time errors. 
The definition we use is that everything on the lower adiabatic surface to the right of the diabatic crossing or on the upper adiabatic surface on the left of the diabatic crossing is the product, and vice versa for the reactant. Mathematically this corresponds to
\begin{equation}
    P_p(t)=h(U_0(t)-U_1(t))\delta_{\activestate(t),-}+h(U_1(t)-U_0(t))\delta_{\activestate(t),+} \label{eq:product_defintion}
\end{equation}
where $h(x)$ is the Heaviside step function and $P_r=1-P_p$. This definition works well for all the cases considered in this work.
We demonstrate numerically in the supporting information that this gives the same rate constants as a purely position-space definition in the normal regime, or a purely adiabatic definition in the inverted regime, while having a shorter transient.

The parameters for the model are taken to be $\beta\Lambda=12$, $\beta\hbar\Omega=1/4$, $\gamma=\Omega$, for a range of values of $\varepsilon$ and $\Delta$. These parameters were chosen to allow a clear comparison of the accuracy of MASH and FSSH, at a reasonable computational cost. In particular the reorganisation energy was chosen to be high enough that the population transfer is in the slow incoherent limit, but low enough that it is possible to run direct population dynamics. This allows us to directly calculate FSSH rates, without needing to employ backward propagation or forward-flux sampling. Additionally, it allows us to demonstrate numerically that in MASH direct population dynamics are equivalent to the results obtained using the flux-correlation formulation, which we show in the supporting information. The characteristic frequency was chosen to make the system as classical as possible without the HEOM calculations becoming too expensive. This was done as our focus here is on assessing the relative accuracy of the dynamics of MASH and FSSH, rather than the importance of nuclear quantum effects. Finally it is known that the effect of overcoherence error becomes less pronounced at high friction,\cite{Falk2014FSSHFriction} and hence to make the test of MASH as stringent as possible we consider a system in the underdamped $\gamma<2\Omega$ regime. Systems with larger reorganisation energy and higher friction are considered in the supporting information.

 %which is high enough to ensure that the population transfer is in the slow incoherent limit, but low enough that it is possible to run direct population dynamics. This allows us to directly calculate FSSH rates, without needing to employ backward propagation or forward flux sampling. Additionally it allows us to demonstrate for MASH that direct population dynamics are both formally and numerically equivalent to the results obtained using the flux-correlation function formulation of MASH in the supporting information. 

\begin{figure}[t]
    \centering
    \includegraphics[width=1.0\linewidth]{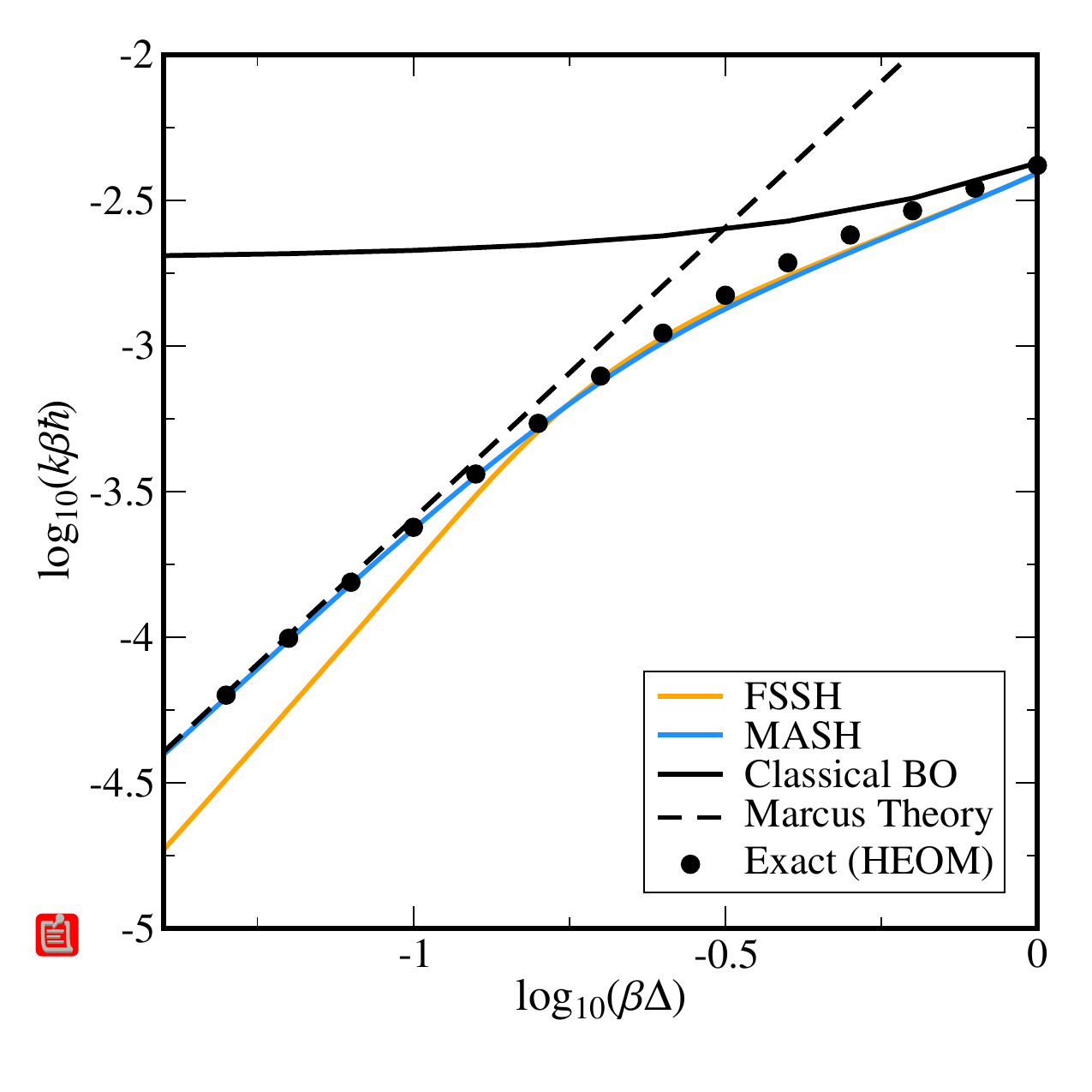}
    \caption{Log-log plot of the rate against the diabatic coupling for a symmetric, $\beta\varepsilon=0$, spin-boson model, with $\beta\hbar\Omega=1/4$, $\gamma=\Omega$ and $\beta\Lambda=12$. FSSH and MASH rates were calculated from the slope of $\langle P_p(t)\rangle$ at the plateau time, between $t=10\beta\hbar$ and $t=20\beta\hbar$.}
    \label{fig:symmetric_short_time_rates_delta_plot}
\end{figure}

\textbf{Results and Discussion:}
Figure~\ref{fig:symmetric_short_time_rates_delta_plot} compares the rates calculated at the plateau time for a symmetric reaction, $\varepsilon=0$, as a function of the diabatic coupling, $\Delta$.
We see that, for intermediate-to-large values of diabatic coupling, $\log_{10}(\beta\Delta)\gtrsim-0.75$,  MASH, FSSH and HEOM all closely agree, 
with the HEOM rate showing only a slight $\sim 10\%$ enhancement due to shallow tunneling. For smaller values of $\Delta$, the reaction approaches the golden-rule regime where Marcus theory is valid. Here we see that MASH continues to closely match the exact results predicted by HEOM, while FSSH begins to deviate significantly with an unphysical slope. This deviation is consistent with previous observations that FSSH struggles in this limit due to its overcoherence error.\cite{Landry2011hopping,Landry2012hopping,Jain2015hopping1,Jain2015hopping2} However, it raises the question: why does MASH not show a similar error? 

\begin{figure}[t]
    \centering
    \includegraphics[width=1.0\linewidth]{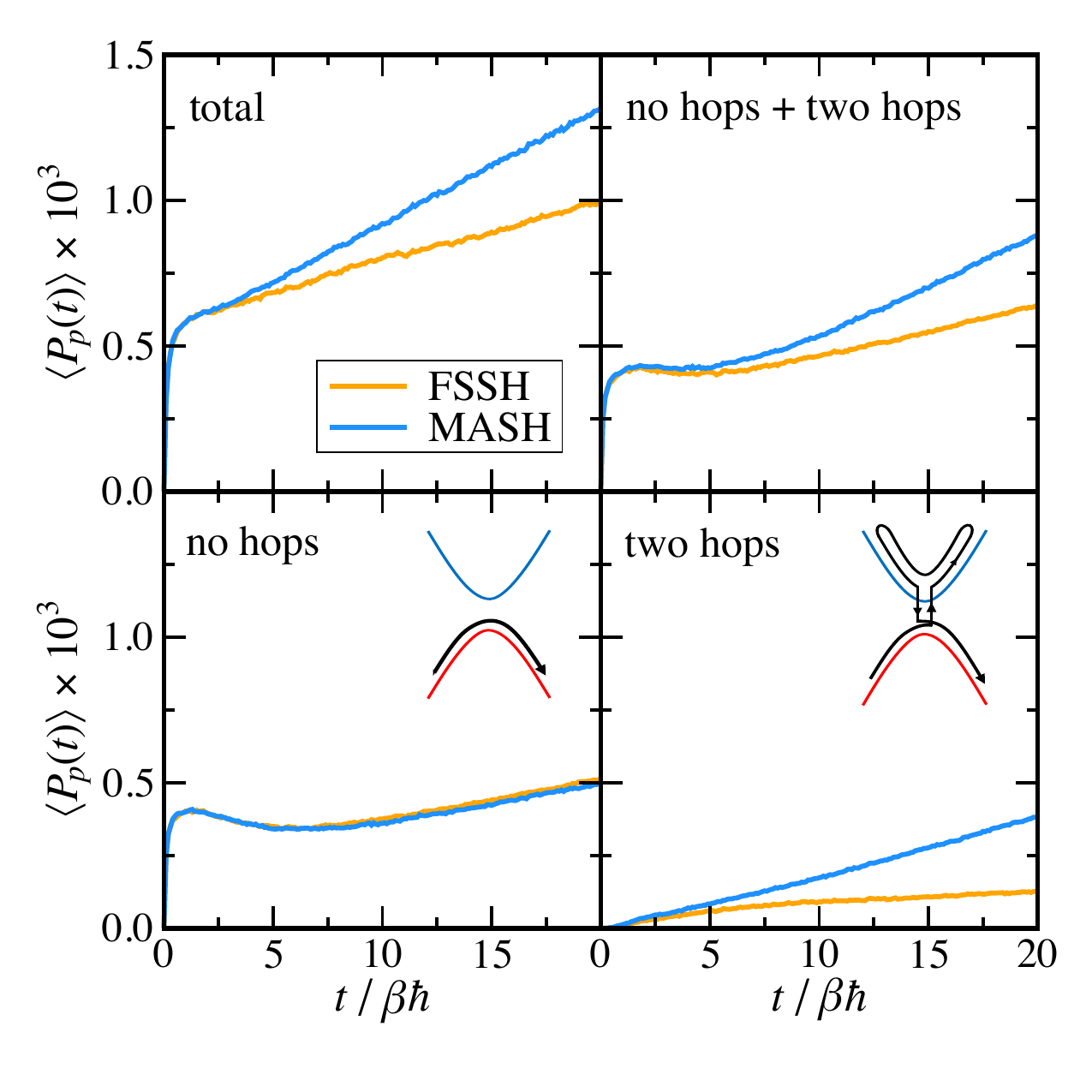}
    \caption{Decomposition of the population of products according to the number of hops for a symmetric, $\beta\varepsilon=0$, spin-boson model, with $\beta\hbar\Omega=1/4$, $\gamma=\Omega$, $\beta\Lambda=12$ in the limit of weak diabatic coupling, $\log_{10}(\beta\Delta)=-7/5$.}
    \label{fig:Population_breakdown}
\end{figure}

To understand this, in Fig.~\ref{fig:Population_breakdown} we analyse $\langle P_p(t)\rangle$ for the smallest value of $\Delta$ considered in Fig.~\ref{fig:symmetric_short_time_rates_delta_plot}. %into contributions from trajectories that have hopped $0$ or $2$ times between $t=0$ and the current time, $t$. % ($\log_{10}(\beta\Delta)=-7/5$).
The top left panel of Fig.~\ref{fig:Population_breakdown} shows the full $\langle P_p(t)\rangle$. Although MASH and FSSH agree during the initial transient, the slope after this time differs significantly, with FSSH predicting a much slower population transfer.
The remaining panels decompose $\langle P_p(t)\rangle$ into contributions from trajectories that have hopped $0$ or $2$ times between $t=0$ and the current time, $t$.
The top right panel shows the sum of the zero- and two-hop trajectories. We see that the difference in the slopes of the MASH and FSSH curves closely resemble those in the full $\langle P_p(t)\rangle$, implying that other terms are contributing only to the transient and not the rate. Hence, to understand the difference between the MASH and FSSH rates, one can focus on just these trajectories.
Unsurprisingly, the no-hop contribution to $\langle P_p(t)\rangle$ (which involves just a single passage through the crossing region) agrees very closely between MASH and FSSH. The key difference occurs in the trajectories that hop twice. The contribution of these trajectories, along with a depiction of a corresponding typical reactive path, is shown in the bottom right panel. From this we see that trajectories that hop twice contribute significantly (and correctly) to the rate in MASH but only contribute a very small amount in FSSH. Hence, the rate predicted by FSSH can be expected to be up to a factor of 2 too small, as previously pointed out by Jain and Subotnik in Ref.~\citenum{Jain2015hopping2}.

\begin{figure}[t]
    \centering
    \includegraphics[width=1.0\linewidth]{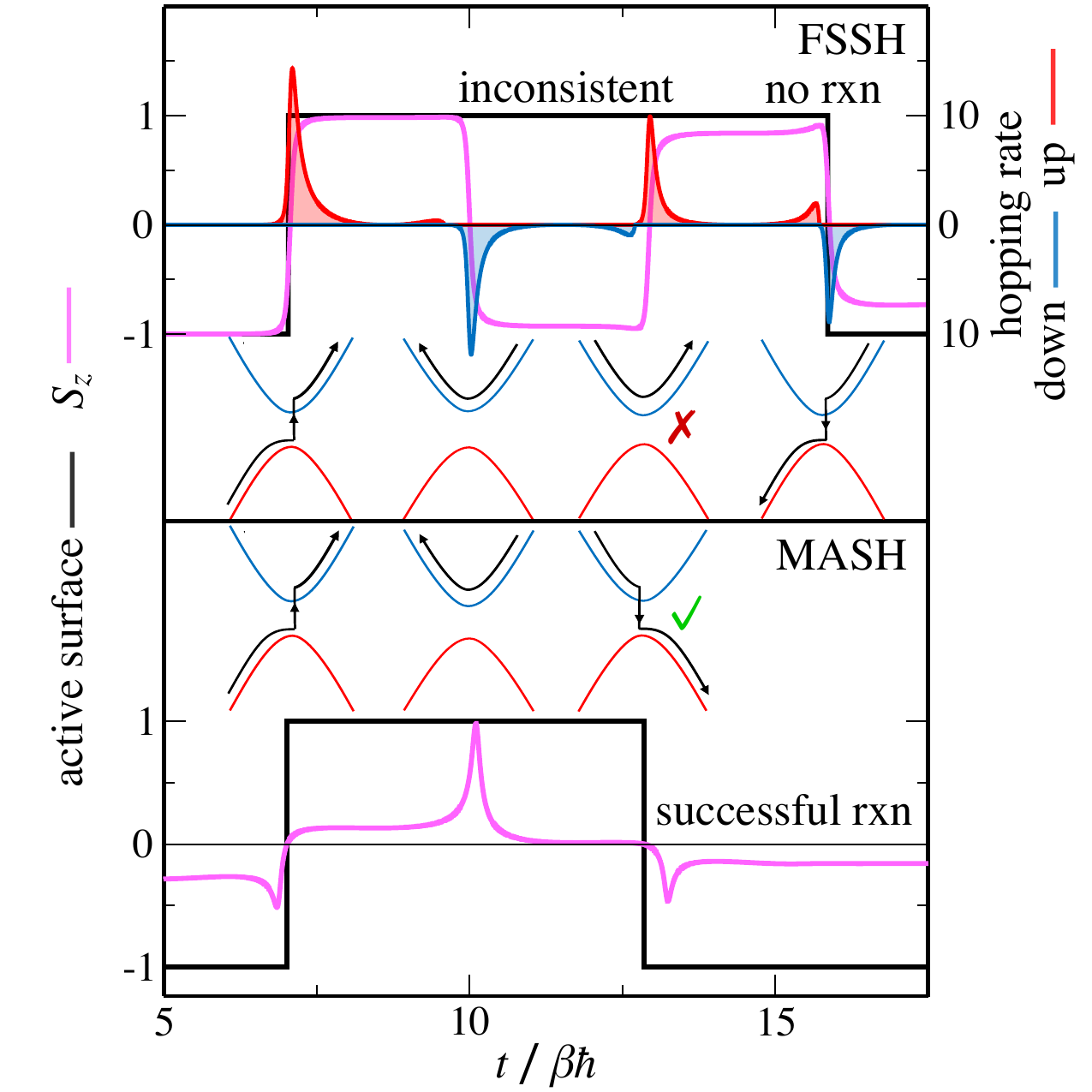}
    \caption{An example of a typical incorrect ``two hop'' FSSH trajectory that fails to react, along with a comparable but correct MASH trajectory. The problem for FSSH occurs on the third crossing, where the wavefunction is inconsistent with the active state. $S_z$ is then predominantly moving up, meaning that the probability to hop down is almost zero. This example is taken from a calculation with   $\beta\hbar\Omega=1/4$, $\gamma=0$, $\beta\varepsilon=0$, $\beta\Lambda=12$.}
    \label{fig:FSSH_inconsistency}
\end{figure}

Having established that it is the two-hop trajectories that differ between MASH and FSSH, it remains to be explained why these trajectories go wrong in FSSH but not in MASH.
Figure~\ref{fig:FSSH_inconsistency} illustrates the behavior of a typical two-hop trajectory in FSSH that ``should'' react but doesn't.
The trajectory starts in the reactant well at $t=0$. 
At $t\approx7\beta\hbar$
 the trajectory reaches the crossing, and hops up due to the strong nonadiabatic coupling and correspondingly large hopping probability. Having hopped up, the trajectory then continues on the upper state before turning around and coming back towards the avoided crossing. Note, at this point the trajectory is not significantly affected by inconsistency or overcoherence error, as the wavefunction coefficients are essentially still in a pure state corresponding to the active surface (i.e.\ $S_z\approx1$). 
At $t\approx10\beta\hbar$
 the trajectory passes through the avoided crossing for a second time and most trajectories hop down (returning to the reactants). However, we follow one of the few that remain on the upper surface (probability $\propto\Delta^2$). Now the wavefunction (which is predominantly in the lower state, $S_z\approx-1$) is inconsistent with the active surface. When the trajectory returns to the avoided crossing for a third time, we expect it to hop down to the product well. However, the wavefunction is evolving in the opposite direction to the expected hop (from down to up instead of up to down). Hence, the probability to hop down is almost zero, and the trajectory incorrectly stays on the upper surface, leading to no reaction. In contrast MASH trajectories cannot have this problem.
When an equivalent MASH trajectory approaches the avoided crossing for the third time, its spin vector is guaranteed to correctly point up (due to the consistency between its spin vector and the active surface).
On passing through the crossing region, its spin vector will then flip down to the lower hemisphere, resulting in a downward hop and a successful reaction.

\begin{figure}[t]
    \centering
    \includegraphics[width=1.0\linewidth]{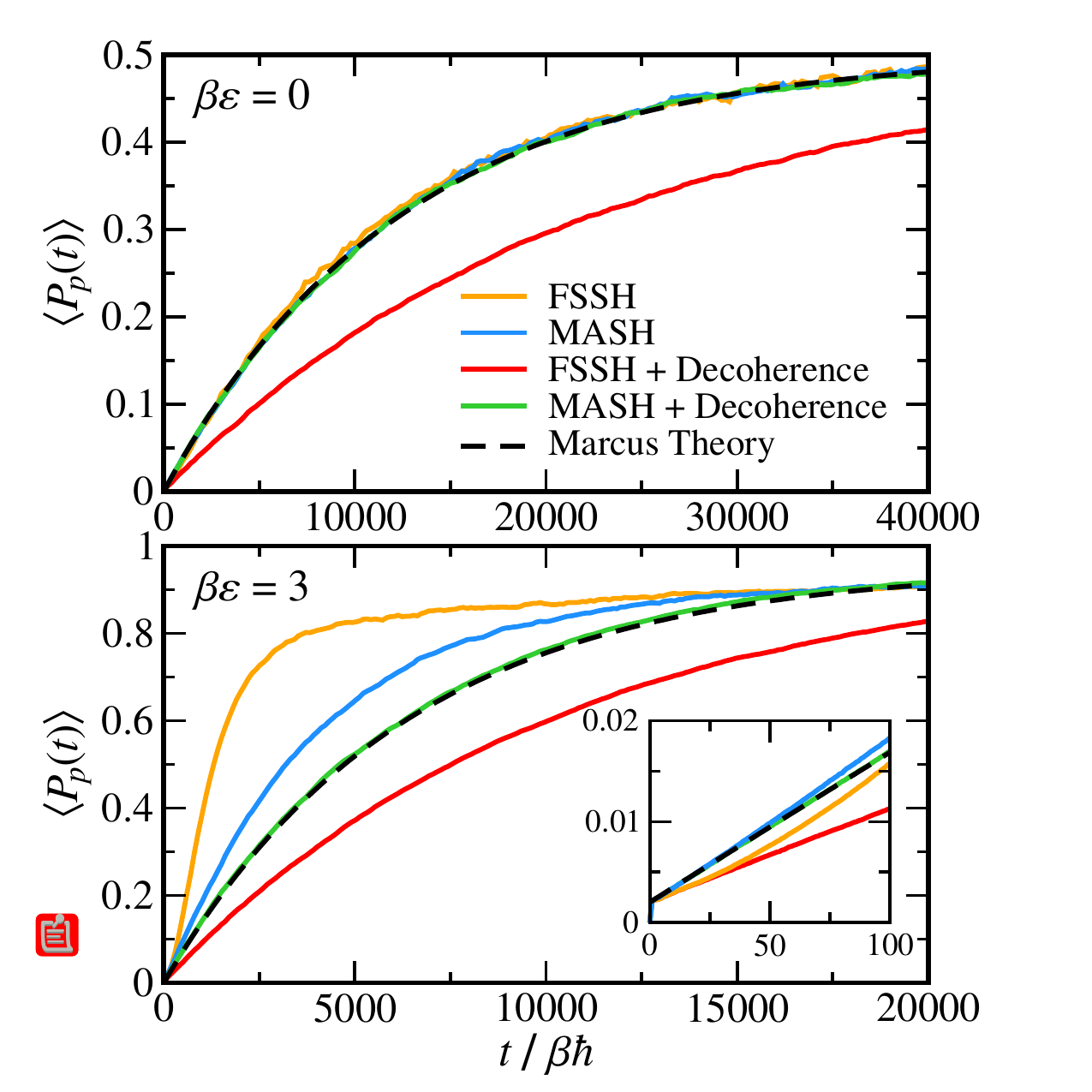}
    \caption{Full population decay for a symmetric, $\beta\varepsilon=0$, and asymmetric, $\beta\varepsilon=3$, spin-boson model, with $\beta\hbar\Omega=1/4$, $\gamma=\Omega$, $\beta\Lambda=12$ in the limit of weak diabatic coupling, $\log_{10}(\beta\Delta)=-7/5$. Inset shows $\langle P_p(t)\rangle$ at short to intermediate time.
Decoherence corrections are applied only when the energy gap is large ($V_+-V_->4 k_{\rm B} T$), making long-time behavior consistent with short to intermediate time. This illustrates that not only is MASH more accurate than FSSH without the application of decoherence corrections, but also that, unlike in FSSH, simple decoherence corrections are sufficient to bring MASH into line with the correct result.
}
    \label{fig:long_time_dependence}
\end{figure}

So far we have only considered the dynamics on the timescale of a single barrier crossing. 
However, in the limit of weak diabatic coupling, the system may come back to the diabatic crossing (region of large nonadiabatic coupling) many times before the reaction takes place. This can lead to a build-up of overcoherence error, causing the long-time rate behavior to deviate significantly from the short-time behavior. 
To investigate this effect, 
Fig.~\ref{fig:long_time_dependence} shows the population of products, $\langle P_p(t)\rangle$, for the full population decay, for two different driving forces, $\beta\varepsilon=0$ and $\beta\varepsilon=3=\beta\Lambda/4$, with all other parameters kept the same as in Fig.~\ref{fig:Population_breakdown}. 

Considering first the upper panel of Fig.~\ref{fig:long_time_dependence}, where ${\beta\varepsilon=0}$, we see immediately that the long-time behavior of both MASH and FSSH agrees perfectly with Marcus theory.
 This is a surprising result,
 as based on the short-time behavior we would expect FSSH to be too slow.
 However, it can be explained away as a fortuitous cancellation of errors due to the symmetry of the model when $\varepsilon=0$. This assumption is confirmed by considering the behavior of an asymmetric reaction, ${\beta\varepsilon=3}$, as shown in the lower panel. The short-time behavior of the symmetric and asymmetric systems are similar as can be seen from the inset.\footnote{See also Fig.~S4 
 of the supporting information.} At long time, however, we see that for the asymmetric system there is no fortuitous cancellation of errors. Instead, the build up of overcoherence error in FSSH leads to a population decay that is noticeably too fast, with a half-life approximately $3.5$ times shorter than Marcus theory. In contrast, MASH goes from being almost exact at short time to being about $1.4$ times too fast at long time.\footnote{It is interesting to note that, despite FSSH and MASH predicting incorrect rates at long time, they nevertheless approach the correct equilibrium populations. This has been observed in previous studies of FSSH,\cite{Schmidt2008equilibrium} and for MASH it was recently proven that under the assumption of ergodicity it is guaranteed to approach the correct long time limit.\cite{thermalizationarxiv}} We see, therefore, that while both MASH and FSSH suffer from a build-up of overcoherence error at long time this error is significantly more pronounced in FSSH. 

The build-up of overcoherence error at long time is of course well established. While it is nice that this error is much smaller in MASH than FSSH, in real simulations on such incredibly long timescales one \emph{should} apply decoherence corrections in both theories. In this regard, the short-time accuracy of MASH also presents a significant advantage. To see why, we note that the application of decoherence corrections is always a balancing act: you need to apply them often enough to fix the overcoherence error, but apply them too often and you will force the system to remain forever on the same adiabat (the quantum Zeno effect). The advantage of MASH is that it requires decoherence corrections less often to obtain accurate results. This means that they can be applied only in regions where it is safe to do so, such as the reactant wells, and not in the vicinity of the coupling region. This makes it more robust and means simpler decoherence schemes can be successfully used. 

In Fig.~\ref{fig:long_time_dependence} we demonstrate this by considering the behavior of MASH and FSSH when one applies a simple decoherence correction. (Note decoherence corrections in MASH correspond to resampling $\bm{S}$ from the hemisphere corresponding to the current active state according to the $|S_z|$ weighting.\cite{Mannouch2023MASH}) For simplicity we use an energy cutoff such that decoherence corrections are applied only when the gap between the states is large, $V_+-V_->4 k_{\rm B} T$, i.e.\ where the system is far from the avoided crossing.\footnote{In this system, this is equivalent to using the more common derivative coupling based cutoff of Ref.~\citenum{Fang1999FSSH_inconsistency}. We choose to give the equivalent energy gap for ease of interpretability.} In the upper panel we see that, for the symmetric system $\varepsilon=0$, the population decay predicted by MASH is unaffected by application of the decoherence correction, leaving it in perfect agreement with Marcus theory. In contrast, however, the FSSH results are made significantly worse by application of the decoherence correction, for reasons explained below. For the asymmetric system, $\beta\varepsilon=3$,  we see that application of the decoherence correction, improves the original MASH result, removing the $\sim40\%$ error, and bringing it into almost perfect agreement with Marcus theory. Again, however, the simple decoherence correction does not fix FSSH, in this case taking the rate from being too fast to too slow.  These results can be understood by noting that by applying the decoherence correction far away from the crossing region we simply make the long-time dynamics consistent with the short-time dynamics. For MASH the short-time dynamics has the correct rate, but for FSSH the short-time dynamics is wrong, as can be seen from the inset, and hence we recover the spuriously low rate seen in Fig.~\ref{fig:symmetric_short_time_rates_delta_plot}. % This highlights the fortuitous nature of the accuracy of the original FSSH results.

That such simple decoherence corrections do not fix FSSH is not a new observation, and for this reason many far more sophisticated decoherence approaches have been developed.\cite{Subotnik2011AFSSH,Jain2016AFSSH} 
However, these methods often come with additional disadvantages, such as increased cost,
and as they are ad-hoc they are not always guaranteed to improve the results.
 The point we would like to stress here is that the increased accuracy of MASH at short times means that decoherence corrections can be applied much more infrequently. For many ultrafast problems this means that they may not be needed at all. But when they are needed they can be  both safer and simpler.

\begin{figure}[t]
    \centering
    \includegraphics[width=1.0\linewidth]{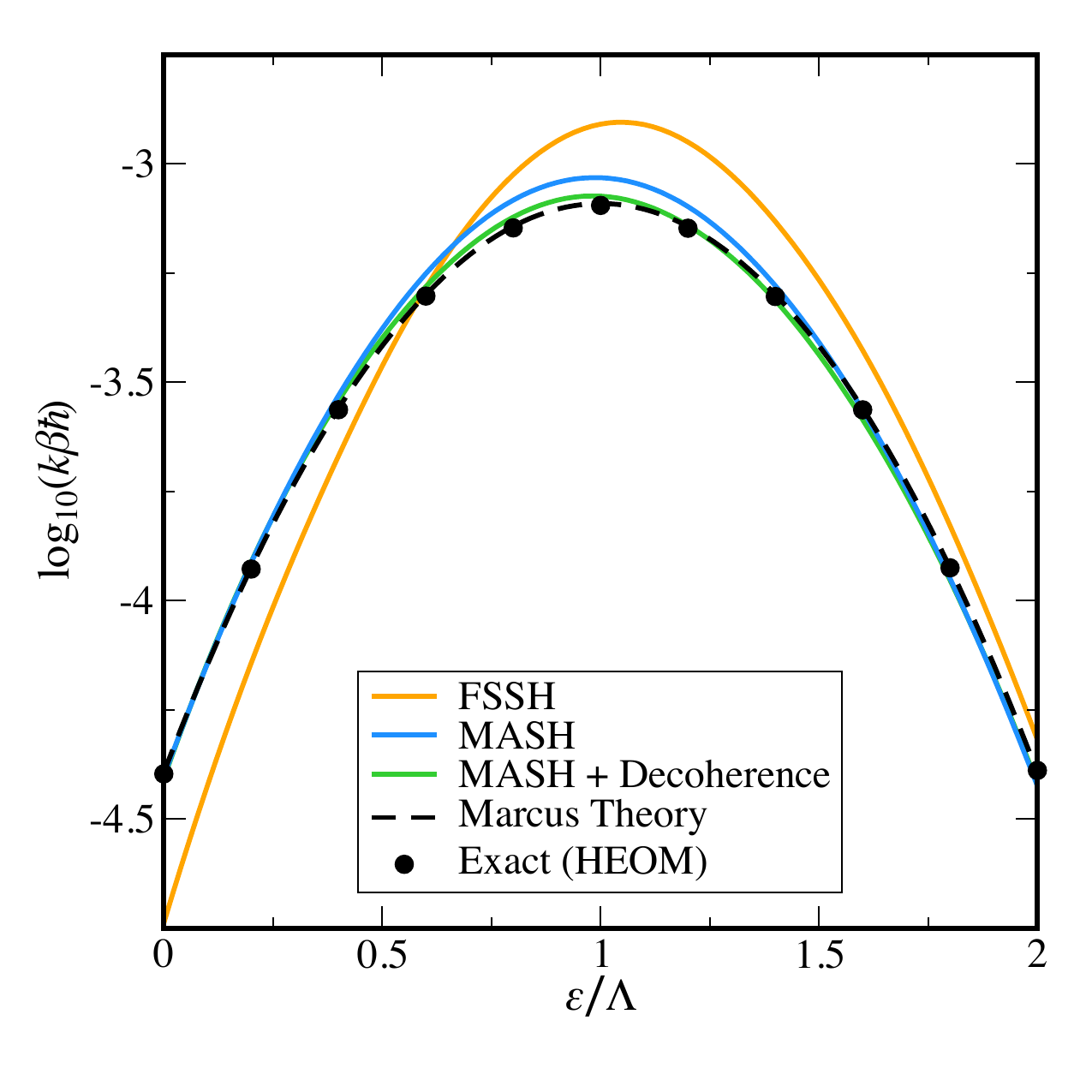}
    \caption{Logarithmic plot of the rate against reaction driving force, showing the famous Marcus turnover behavior for a spin-boson model with weak diabatic coupling, $\log_{10}(\beta\Delta)=-7/5$, $\beta\hbar\Omega=1/4$, $\gamma=\Omega$, and $\beta\Lambda=12$. FSSH and MASH rates were calculated from the slope of $\langle P_p(t)\rangle$, at the plateau time, between $t=10\beta\hbar$ and $t=20\beta\hbar$.}
    \label{fig:inverted_regime_plot}
\end{figure}

Finally, having understood the difference between MASH and FSSH, we consider % the ability of MASH to recover
the famous Marcus turnover curve. Figure~\ref{fig:inverted_regime_plot} shows the behavior of the rate in the weak-coupling limit ($\log_{10}(\beta\Delta)=-7/5$) as a function of the bias to products, $\varepsilon$.
As in Fig.~\ref{fig:symmetric_short_time_rates_delta_plot} the FSSH and MASH rates are calculated from the slope of $\langle P_p(t)\rangle$ near the plateau time, between $t=10\beta\hbar$ and $t=20\beta\hbar$. As expected from the results above, FSSH deviates significantly from Marcus theory and the exact results, showing an unphysical asymmetry about $\varepsilon=\Lambda$. In contrast MASH reproduces both the exact results and Marcus theory very well. MASH is also not perfectly symmetric due to a slightly larger error deep in the inverted regime ($\varepsilon=2\Lambda$) than in the symmetric case ($\varepsilon=0$). However, in both cases the errors are less than $10\%$. The largest error in MASH is observed close to the activationless limit $\varepsilon/\Lambda=1$. Here the MASH rate is about $15\%$ higher than the Marcus-theory result. In contrast FSSH is about $60\%$ too large. As the avoided crossing is located at the minimum of the reactant well in the activationless case, this leads to a faster build up of overcoherence error and the increase in the rate seen at long time in the lower panel of Fig.~\ref{fig:long_time_dependence} starts to affect the dynamics even at the short times considered here.
This is confirmed by application of the same decoherence correction as was used in Fig.~\ref{fig:long_time_dependence}, which stops the build up of overcoherence error, resulting in MASH rates that are within $7\%$ of the exact rate for the full range of $\varepsilon$ considered. As in Fig.~\ref{fig:long_time_dependence} the increased error of FSSH at short time means that this simple decoherence correction is not sufficient to fix the inconsistency error of FSSH, and hence the rates still deviate significantly from the Marcus-theory result, as can be seen in Fig.~S6 
 of the supporting information.

\textbf{Conclusions:} 
It is well established that the overcoherence error of FSSH is most pronounced for the calculation of reaction rates in the limit of weak diabatic coupling (the Marcus-theory regime).\cite{Landry2011hopping,Landry2012hopping,Jain2015hopping1,Jain2015hopping2,Falk2014FSSHFriction} Here we have revisited this problem to assess the accuracy of a newly proposed alternative to FSSH, the mapping approach to surface hopping (MASH). In comparing MASH and FSSH, we have considered two different timescales: the timescale of a single barrier-crossing event, $t_{\rm pl}$, and the timescale of the reaction, $\tau_{\rm rxn}$. 

On the timescale of barrier crossing, MASH provides a significant improvement upon FSSH, accurately recovering the results of Marcus theory without the use of decoherence corrections.
This might seem surprising at first, as it is not immediately obvious how MASH, which is also an independent trajectory method, 
is able to capture decoherence. However, we have shown that the improvement can be explained in terms of the dramatic inconsistency between the active state and wavefunction coefficients, which can exist in FSSH but is absent from MASH.
 %which exists in FSSH but is absent from MASH. 
 
On very long timescales, MASH again provides a significant improvement over FSSH.  While overcoherence error does still build up in MASH, we have found it to be much less significant than in FSSH.  This can again be explained in terms of the inconsistency in FSSH, which means that the build-up of error can be sudden and large, whereas in MASH the build-up of error is more gradual and ultimately smaller. Perhaps most importantly, the increased accuracy of MASH over FSSH at short time means that, when they are used, decoherence corrections need only be applied well away from the coupling region, making them safer and simpler to use.

 MASH also comes with additional practical advantages over FSSH in the calculation of reaction rates. In particular, as the dynamics of MASH are deterministic and obey time-translation symmetry, there is no need for approximate backwards-time propagation or advanced methods such as forward-flux sampling. One can instead rigorously apply the flux-correlation formalism and related techniques, such as the Bennett--Chandler method, in order to efficiently calculate reaction rates.
Given these significant improvements, and the fact that MASH is simple to use, requires only relatively minor modifications to existing FSSH code, and can be run at equivalent computational cost, MASH has the potential to replace FSSH as the go-to method for the simulation of nonadiabatic processes.

The only thing limiting MASH as a replacement to FSSH is that the current theory is restricted to two-state problems. Recently a modification to MASH has been proposed, designed for application to multistate problems.\cite{Runeson2023MASH} However this theory is a different method to the MASH described here. It does not reduce to the current theory in the case of a two-level system, and although it is accurate for many problems, it was shown to be significantly less accurate for the timescales of population decay in a spin-boson model in the inverted regime.  Work to develop a multistate generalisation of the present MASH method is in progress, and if this can be achieved, while retaining the advantages of the two-state theory, it would present a significant challenge to the hegemony of FSSH.% in the study of electronically nonadiabatic chemistry.

Finally, we note that we have focused here exclusively on the limit of classical nuclei. It is, however, well known that nuclear quantum effects, in particular tunneling and zero-point energy, can have a significant effect on the rate of nonadiabatic reactions, such as electron transfer, intersystem crossing and proton coupled electron transfer.\cite{ChandlerET,Oxygen,CIinst,Blumberger2008ET,Layfield2014PCET_SLO,Lawrence2020FeIIFeIII,Heller2021Thiophosgene,Heller2022Nitrenes} In recent years there has been a continued interest in the development of methods that can accurately incorporate nuclear quantum effects in the simulation of electronically nonadiabatic reactions.\cite{Shushkov2012RPSH,mapping,Ananth2013MVRPMD,Chowdhury2017CSRPMD,Tao2018isomorphic,Tao2019RPSH,Sindhu2021NQE_FSSH} While there has been significant development in methods specialised for accurately predicting thermal reaction rates,\cite{GoldenGreens,inverted,Lawrence2020Improved,Lawrence2018Wolynes,Lawrence2020rates,Trenins20224thOrder} at present there is no fully dynamical method that can offer comparable accuracy.\cite{Lawrence2019isoRPMD}  This is in part due to the difficulty that such dynamical methods face in accurately describing rates even in the limit of classical nuclei. In this regard, the results of the present study indicate that MASH provides a new and exciting route to the development of a fully dynamical nonadiabatic theory capable of accurately describing nuclear tunneling and zero-point energy.

\rev{
\textbf{Supporting Information:}
Details on rate calculations
and additional results to support the conclusions of the paper.
}

\begin{acknowledgement}
JEL was supported by an ETH Zurich Postdoctoral Fellowship and JRM was supported by the Cluster of Excellence ``CUI: Advanced Imaging of Matter'' of the Deutsche Forschungsgemeinschaft (DFG) – EXC 2056 – project ID 390715994.
\end{acknowledgement}

\bibliography{references,extra_refs}

\end{document}

\maketitle

\renewcommand{\thepage}{S\arabic{page}}
\renewcommand{\theequation}{S\arabic{equation}}
\renewcommand{\thefigure}{S\arabic{figure}}
\renewcommand{\thetable}{S\arabic{table}}
\renewcommand{\thesection}{S\arabic{section}}
\renewcommand{\thesubsection}{S\arabic{section}.\arabic{subsection}}

\section{Key Definitions.}
To simplify notation, it is helpful to define the trace (or phase-space integral) over the MASH variables as 
\begin{equation}
    \tr[A(\bm{p},\bm{q},\bm{S})] = \frac{1}{(2\pi\hbar)^f} \int \mathrm{d}\bm{q} \int \mathrm{d}\bm{p} \int \mathrm{d} \bm{S} A(\bm{p},\bm{q},\bm{S}),
\end{equation}
where $f$ is the number of nuclear degrees of freedom. The integral over $\bm{S}$ is defined as
\begin{equation}
    \int \mathrm{d} \bm{S} A(\bm{S}) = \frac{1}{2\pi}\int_0^\pi \mathrm{d} \theta \int_0^{2\pi} \mathrm{d} \phi \sin(\theta) A(\bm{S}),
\end{equation}
with 
\begin{subequations}
\begin{equation}
     S_x = \sin(\theta)\cos(\phi)   
\end{equation}
\begin{equation}
     S_y = \sin(\theta)\sin(\phi)   
\end{equation}
\begin{equation}
     S_z = \cos(\theta).   
\end{equation}
\end{subequations}
The trace over the FSSH variables can be defined similarly, the key difference being that, as the active surface, $\activestate$, is not obtained deterministically from the Bloch sphere, it must be introduced as a separate variable. Hence, we define the FSSH trace as
\begin{equation}
    \tr[A(\bm{p},\bm{q},\bm{S},\activestate)] = \frac{1}{(2\pi\hbar)^f} \sum_{\activestate=\pm} \int \mathrm{d}\bm{q} \int \mathrm{d}\bm{p} \int \mathrm{d} \bm{S} A(\bm{p},\bm{q},\bm{S},\activestate) .
\end{equation}

It will also be helpful to define the MASH and FSSH total energy as
\begin{equation}
E(\bm{p},\bm{q},\activestate) = H_+(\bm{p},\bm{q})\delta_{+,\activestate}+H_-(\bm{p},\bm{q})\delta_{-,\activestate},
\end{equation}
where
\begin{equation}
  H_\pm(\bm{p},\bm{q}) = \sum_{j=1}^f \frac{p_j^2}{2m_j} + V_\pm(\bm{q}),
\end{equation}
are the energies on each of the adiabatic states, and we remind the reader that in MASH
\begin{equation}
  \activestate(\bm{S}) = \mathrm{sign}(S_z).
\end{equation}

\section{Numerical calculation of MASH and FSSH rates.}
For a system which undergoes incoherent dynamics at long time, the rate constant to pass from reactants to products can be written formally as \cite{Craig2007condensed,Lawrence2019ET}
\begin{equation}
    k = \lim_{t\to\infty} \frac{\frac{\mathrm{d}}{\mathrm{d}t}\langle{P}_p(t)\rangle}{1-\langle P_p(t)\rangle/\langle P_p(\infty)\rangle}.
\end{equation}
Often this is simplified by introducing the idea of the plateau time, $t_{\rm pl}$,\cite{ChandlerGreen} which is short enough that $\langle P_p(t)\rangle/\langle P_p(\infty)\rangle\approx0$, but long enough that the dynamics has settled into the exponential decay, such that
\begin{equation}
    k\approx \frac{\mathrm{d}}{\mathrm{d}t}\langle{P}_p(t)\rangle \bigg|_{t=t_{\rm pl}}.
\end{equation}

For the results presented in the main paper, we compute the MASH and FSSH rates by taking the average of  
\begin{equation}
    k(t) = \frac{\frac{\mathrm{d}}{\mathrm{d}t}\langle{P}_p(t)\rangle}{1-\langle P_p(t)\rangle/\langle P_p(\infty)\rangle}
\end{equation}
between $t=10\beta\hbar$ and $t=20\beta\hbar$, where the derivative is computed by finite difference from $\langle{P}_p(t)\rangle$.  In MASH this is defined as
\begin{equation}
\label{eq:prod_MASH}
\begin{aligned}
    \langle P_p(t) \rangle =  \frac{\tr\left[e^{-\beta E} |S_z| P_r  P_p(t)\right]}{\tr\left[e^{-\beta E} |S_z| P_r \right]},
\end{aligned}
\end{equation}
and in FSSH as
\begin{equation}
\begin{aligned}
    \langle P_p(t) \rangle =  \frac{\tr\left[e^{-\beta E} \delta(S_z-\activestate)  P_r P_p(t)\right]}{\tr\left[e^{-\beta E} \delta(S_z-\activestate) P_r \right]}.
\end{aligned}
\end{equation}
As stated in the body of the main paper, in both MASH and FSSH we use the following general definition of the reactants and products, valid in both the normal and inverted Marcus regimes,
\begin{subequations}
\label{Product_reactant_defintions}
\begin{equation}
    P^{(g)}_p(t)=h(U_0(t)-U_1(t))\delta_{\activestate(t),-}+h(U_1(t)-U_0(t))\delta_{\activestate(t),+}
\end{equation}
\begin{equation}
    P^{(g)}_r(t)=h(U_0(t)-U_1(t))\delta_{\activestate(t),+}+h(U_1(t)-U_0(t))\delta_{\activestate(t),-}, 
\end{equation}
\end{subequations}
where $h(x)$ is the Heaviside step function. Here we have introduced an additional superscript $(g)$ to distinguish this general definition from the position, and adiabatic definitions that are not used in the main paper, but are considered in this supporting information.   We remind the reader that  $P^{(g)}_r = 1-P^{(g)}_p$.

\subsection{Equivalence of different definitions of reactants and products.}
\begin{figure}[t]
    \centering
    \includegraphics[width=0.5\linewidth]{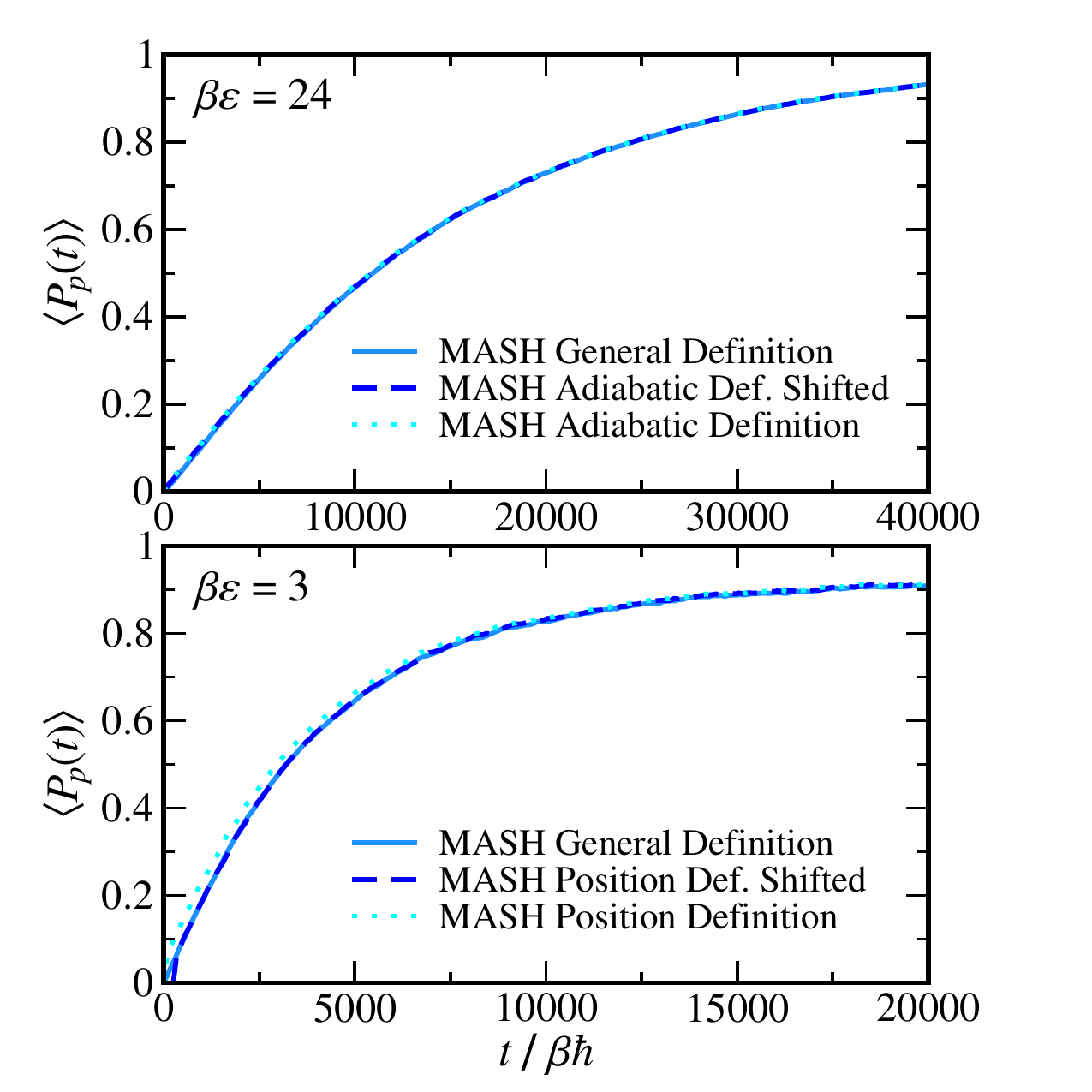}
    \caption{Comparison of $\langle P_p(t)\rangle$ for different definitions of the reactants and products at long time. Upper panel: inverted regime $\beta\varepsilon=24$, with $\log_{10}(\beta\Delta)=-7/5$, $\beta\hbar\Omega=1/4$, $\gamma=\Omega$ and $\beta\Lambda=12$. The dashed curve is shifted by $t=60\beta\hbar$ to better illustrate the equivalence with the general definition at long time. Lower panel: normal regime $\beta\varepsilon=3$, with $\log_{10}(\beta\Delta)=-7/5$, $\beta\hbar\Omega=1/4$, $\gamma=\Omega$ and $\beta\Lambda=12$. The dashed curve is shifted by $t=250\beta\hbar$ to better illustrate the equivalence with the general definition at long time.}
    \label{Fig:SI_dividing_surface_comparison}
\end{figure}
Here we demonstrate numerically that the definition of reactants and products given in Eq.~\ref{Product_reactant_defintions} give rates which are identical to more common definitions of reactants and products. Firstly in the normal regime, we compare to a purely position space definition of the reactants and products
\begin{subequations}
\label{eq:position_defintion}
\begin{equation}
    P^{(q)}_p(t) = h(U_0(t)-U_1(t))
\end{equation}
\begin{equation}
   P^{(q)}_r(t) = h(U_1(t)-U_0(t)).   
\end{equation}
\end{subequations}
Secondly, in the inverted regime we compare to using a purely adiabatic state definition of the reactants and products
\begin{subequations}
\label{eq:adiabatic_defintion}
\begin{equation}
    P^{(a)}_p(t) = \delta_{\activestate(t),-}
\end{equation}
\begin{equation}
   P^{(a)}_r(t) = \delta_{\activestate(t),+}.   
\end{equation}
\end{subequations}
Figure~\ref{Fig:SI_dividing_surface_comparison} compares $\langle P_p(t)\rangle$ calculated using the general definition (used in the main paper) against the position and adiabatic definitions. The lower panel shows the results for a system in the normal regime and the upper panel shows the results for a system in the inverted regime. We see in both cases that at long time the definition used in the main paper is equivalent to the more commonly used definitions given in Eqs.~\ref{eq:position_defintion} and \ref{eq:adiabatic_defintion}. 

\section{Flux-Correlation version of MASH.}
It is typically more efficient to calculate $\frac{\mathrm{d}}{\mathrm{d}t}\langle P_p(t)\rangle$ using the flux-correlation formalism, rather than from a direct simulation of $\langle P_p(t)\rangle$.\cite{ChandlerGreen} To arrive at the flux-correlation formalism, one differentiates analytically and then shifts the time origin. As discussed in the main paper, it is not possible to do this within FSSH\@.  However it is possible for MASH\@. The following demonstrates that this is the case and derives the resulting flux-correlation MASH expressions.

\subsection{Time symmetries.}
MASH dynamics obeys a number of important properties that allow one to use the flux-correlation formalism to efficiently calculate reaction rates. 
\subsubsection{Time translation.}  
The first of these properties is time-translation symmetry. As the dynamics has no explicit time dependence, we can begin by simply changing the origin of time to give
\begin{equation}
\begin{aligned}
\tr[e^{-\beta E(\bm{p}_0,\bm{q}_0,\bm{S}_0)} A(0)B(t)]&=
    \frac{1}{(2\pi\hbar)^f} \int \mathrm{d}\bm{q}_0 \int \mathrm{d}\bm{p}_0 \int \mathrm{d} \bm{S}_0 \, e^{-\beta E(\bm{p}_0,\bm{q}_0,\bm{S}_0)} A(0)B(t) \\
    &=\frac{1}{(2\pi\hbar)^f} \int \mathrm{d}\bm{q}_{-t} \int \mathrm{d}\bm{p}_{-t} \int \mathrm{d} \bm{S}_{-t} \, e^{-\beta E(\bm{p}_{-t},\bm{q}_{-t},\bm{S}_{-t})} A(-t)B(0).
\end{aligned}
\end{equation}
Now as proved in the original MASH paper (Ref.~\citenum{Mannouch2023MASH}) the MASH dynamics conserve both the measure and the energy, hence 
\begin{equation}
\begin{aligned}
\tr[e^{-\beta E(\bm{p}_0,\bm{q}_0,\bm{S}_0)} A(0)B(t)]
    &=\frac{1}{(2\pi\hbar)^f} \int \mathrm{d}\bm{q}_{0} \int \mathrm{d}\bm{p}_{0} \int \mathrm{d} \bm{S}_{0} \, e^{-\beta E(\bm{p}_{0},\bm{q}_{0},\bm{S}_{0})} A(-t)B(0)\\
    &=\tr[e^{-\beta E(\bm{p}_0,\bm{q}_0,\bm{S}_0)} A(-t)B(0)].
\end{aligned}
\end{equation}

It is important to note that, as discussed in the original MASH paper,\cite{Mannouch2023MASH} MASH does not exactly obey detailed balance for the calculation of reaction rates. This is because the dynamics does not in general conserve $|S_z(t)|$ and hence
\begin{equation}
\begin{aligned}
   \tr\left[e^{-\beta E} |S_z| P_r  P_p(t)\right] \neq \tr\left[e^{-\beta E} |S_z| P_p P_r(-t) \right].
\end{aligned}
\end{equation}
Therefore while the dynamics does obey time-translational symmetry, the correlation functions do not when $|S_z|$ is treated as part of the initial distribution. However, as we shall explain below, the fact that the dynamics obey time-translation symmetry is enough to develop an efficient flux-correlation formalism. Finally we note that although MASH does not formally obey detailed balance, under the assumption of ergodicity, MASH is guaranteed to correctly thermalise in the long-time limit.\cite{thermalizationarxiv} Hence MASH is accurate both at short time and at long time.
 
\subsubsection{Time-inversion symmetry.} The MASH equations of motion are given, in the adiabatic basis, by \begin{subequations}
		\label{eq:mash_eom}
		\begin{align}
			\dot{S}_{x}&=\sum_{j}\frac{2d_{j}(\bm{q})p_{j}}{m_j}S_{z}-\frac{V_{+}(\bm{q})-V_{-}(\bm{q})}{\hbar}S_{y}  \\
			\dot{S}_{y}&=\frac{V_{+}(\bm{q})-V_{-}(\bm{q})}{\hbar}S_{x}  \\
			\dot{S}_{z}&=-\sum_{j}\frac{2d_{j}(\bm{q})p_{j}}{m_j}S_{x}   \\
			\dot{q_{j}}&=\frac{p_{j}}{m_j} \\
			\dot{p}_{j}&=-\frac{\partial V_{+}(\bm{q})}{\partial q_{j}}h(S_{z})-\frac{\partial V_{-}(\bm{q})}{\partial q_{j}}h(-S_{z})+2[V_{+}(\bm{q})-V_{-}(\bm{q})]d_{j}(\bm{q})S_{x}\delta(S_{z}).
		\end{align}
	\end{subequations}
These equations of motion are left unchanged under the transformation: $t\mapsto-t$, $S_{y}\mapsto-S_{y}$ and $p_{j}\mapsto-p_{j}$. This means than any expression involving backward-propagated trajectories can be converted into an expression involving forward-propagated trajectories, by reversing the signs of $S_y$ and $p_j$.

\subsection{Flux-correlation function.}
Using the time-translation symmetry, we can straightforwardly obtain the following general expression for the product population derivative
\begin{equation}
\begin{aligned}
    \frac{\mathrm{d}}{\mathrm{d}t}\langle {P}_p(t) \rangle = & \frac{\tr\left[e^{-\beta E} \dot{P}_p(0) P_r(-t) |S_z(-t)| \right]}{\tr\left[e^{-\beta E}P_r |S_z|\right]}. \label{eq:general_expectation_of_P_p_dot}
\end{aligned}
\end{equation}
Before we can use the time-inversion symmetry to  simplify further, we need to specialise to the particular definition of the reactants and products. In the following, we consider first the MASH flux-correlation formalism when the reactants and products are defined in terms of position or adiabatic states, as in Eqs.~\ref{eq:position_defintion} and \ref{eq:adiabatic_defintion}. Then we will consider the flux-correlation formalism for the general definition of reactants and products used in the main paper (Eq.~\ref{Product_reactant_defintions}), which combines results from both the position and adiabatic definitions. In each case, we need to evaluate the derivative, $\dot{P}_p(0)$.

\subsubsection{Position based definition (normal regime).}
We begin with considering a position based definition. To do so, it is helpful to simplify notation by defining the generalised coordinate, $Q$, such that $Q=Q^\ddagger$ corresponds to $U_0(\bm{q})=U_1(\bm{q})$. With this definition, we have that
\begin{subequations}
\begin{equation}
    P^{(q)}_r(t) = h(-(Q(t)-Q^\ddagger))
\end{equation}
\begin{equation}
   P^{(q)}_p(t) = h(Q(t)-Q^\ddagger).   
\end{equation}
\end{subequations}
 Hence, evaluating the time derivative of $P^{(q)}_p(t)$, we obtain
\begin{equation}
   \dot{P}^{(q)}_p(t) = \dot{Q}(t)\delta(Q(t)-Q^\ddagger).  \label{eq:position_derivative} 
\end{equation}
Inserting this into Eq.~\ref{eq:general_expectation_of_P_p_dot} and employing the time-inversion symmetry, we obtain 
\begin{equation}
\begin{aligned}
    \langle \dot{P}^{(q)}_p(t) \rangle = & -\frac{\tr\left[e^{-\beta E}\dot{{Q}}\delta({Q}-Q^\ddagger) |{S}_z(t)| {P}^{(q)}_r(t)\right]}{\tr\left[e^{-\beta E}P^{(q)}_r |S_z|\right]}. \label{eq:position_side_flux}
\end{aligned}
\end{equation}
Except for the MASH weighting factor, $|S_z(t)|$, this has an identical form to the standard approach for calculating classical rates within the Born--Oppenheimer approximation.\cite{ChandlerGreen} We note that by considering the equivalent expression calculated with the MASH weighting factor at the initial time, $|S_z(0)|$, one obtains a very useful estimate of the MASH error.\cite{Mannouch2023MASH} 

\subsubsection{Adiabatic definition (inverted regime).}
To evaluate the derivative of $P_p(t)$ in Eq.~\ref{eq:adiabatic_defintion}, it is helpful to first use the definition of the active state in MASH to rewrite the reactant and products as
\begin{subequations}
    \begin{equation}
    P^{(a)}_r(t) = h(S_z(t))
\end{equation}
\begin{equation}
    P^{(a)}_p(t) = h(-S_z(t))=1-h(S_z(t)).
\end{equation}
\end{subequations}
Evaluating the derivative requires care, as the MASH dynamics is not analytic at $t_{\rm hop}$ (where $S_z(t_{\rm hop})=0$) and hence
\begin{equation}
    \frac{\mathrm{d}}{\mathrm{d}t}h(S_z(t))\neq \dot{S}_z(t)\delta(S_z(t)).
\end{equation}
We therefore return to the definition of the derivative
\begin{equation}
  \frac{\mathrm{d}}{\mathrm{d}t} h(S_z(t)) = \lim_{\epsilon\to0} \frac{h(S_z(t+\epsilon))-h(S_z(t))}{\epsilon},
\end{equation}
and carefully take the limit such that $t$ is not exactly the hopping time, but rather (for $\epsilon\to0^+$) infinitesimally before or (for $\epsilon\to0^-$) infinitesimally after the hop. To evaluate the derivative we thus need to consider the behaviour of $h(S_z(t+\epsilon))-h(S_z(t))$ for small values of $\epsilon$. It is clear that this can only be non-zero if there is a change in state between the time $t$ and $t+\epsilon$. Assuming that $\epsilon$ is small enough that there is at most one attempted hop (where $S_z=0$) between $t$ and $t+\epsilon$, a careful consideration of the possibilities (explained in detail below) results in the equation
\begin{equation}
\begin{aligned}
    &h(S_z(t+\epsilon))-h(S_z(t)) =\left[h(S_z(t)+\epsilon\dot{S}_z(t))-h(S_z(t))\right]\Big[h(S_z(t))+h(-S_z(t))\,h(\mathit{\Delta} E_{\rm hop}(t))\Big], \label{short_time_population_change}
    \end{aligned}
\end{equation}
where $\mathit{\Delta} E_{\rm hop}$ is the difference between the kinetic energy along the derivative coupling vector and the adiabatic energy gap
\begin{equation}
    \mathit{\Delta} E_{\rm hop} = \frac{1}{2} \frac{(\tilde{\bm{p}} \cdot \tilde{\bm{d}})^2}{\tilde{\bm{d}}\cdot\tilde{\bm{d}}}  - \left(V_+(\bm{q}) - V_-(\bm{q}) \right),
\end{equation}
where $\tilde{p}_j=p_j/\sqrt{m_j}$ and $\tilde{d}_j=d_j/\sqrt{m_j}$ are the mass-weighted momentum and derivative coupling vectors respectively.
To understand Eq.~\ref{short_time_population_change}, we note first that for the left-hand side to be non-zero the active state must change between $t$ and $t+\epsilon$.
The first of the two terms on the right-hand side of Eq.~\ref{short_time_population_change} corresponds to whether a hop is attempted and the second to whether it has enough energy to actually occur. The first term is only non-zero if the current rate of change predicts a change in the sign of $S_z$ between $t$ and $t+\epsilon$ (i.e.~an attempted hop). The second term then accounts for the possibility that the hop is frustrated, in which case the overall expression must be zero (due to the reflection of $S_{z}$ off the spin-sphere equator for a frustrated hop). To see that this is the correct expression, note that if the system is on the upper surface, $h(S_z(t))=1$, then the hop is necessarily not frustrated (this is true whether $t$ is just before or just after a hop, i.e.\ whether $\epsilon$ is positive or negative). However, if the system is on the lower surface, $h(-S_z(t))=1$, then it will hop (or just have hopped) if and only if the kinetic energy along the derivative coupling vector is greater than the adiabatic energy gap, $\mathit{\Delta}E_{\rm hop}(t)>0$. 

As we have already alluded to, there is a choice in whether the limit is taken from above ($\epsilon\to0^{+}$) or from below ($\epsilon\to0^{-}$). First we will take the limit as $\epsilon\to0^{-}$. This will result in expressions for which $t$ is infinitesimally after the hopping time, rather than before.
Noting that only the first term on the right-hand side of Eq.~\ref{short_time_population_change} involves $\epsilon$, the key to evaluating the limit is the following identity
\begin{equation}
\begin{aligned}
    \lim_{\epsilon\to0^{-}}\frac{h(x+\epsilon b)-h(x)}{\epsilon}
    %&=\lim_{\epsilon\to0^{-}}|b|\frac{h(x+\epsilon b)-h(x)}{\epsilon|b|}\\
    %&=\lim_{\epsilon\to0^{-}}h(b)|b|\frac{h(x+\epsilon )-h(x)}{\epsilon}\\&+\lim_{\epsilon\to0^{-}}h(-b)|b|\frac{h(x-\epsilon )-h(x)}{\epsilon}\\
    &=\lim_{\eta\to0^{+}}\delta(x-\eta)h(b)b+\lim_{\eta\to0^{-}}\delta(x-\eta)h(-b)b\\
    &=\delta_+(x)h(b)b+\delta_-(x)h(-b)b,\label{one_sided_deltafns}
    \end{aligned}
\end{equation}
where we will replace $x$ and $b$ with $S_z(t)$ and $\dot{S}_z(t)$. This result is straightforward to verify by considering the two cases $b>0$ and $b<0$ separately.
 Combining this with Eq.~\ref{short_time_population_change}, and utilising trivial identities such as $\delta_{+}(S_z(t))h(-S_z(t))=0$, one then obtains the following expression for the time derivative 
\begin{equation}
    \frac{\mathrm{d}}{\mathrm{d}t} h(S_z(t))=\delta_+(S_z(t))\,h(\dot{S}_z(t))\dot{S}_z(t)
    +\delta_-(S_z(t))\,h(-\dot{S}_z(t))\dot{S}_z(t) \,h(\mathit{\Delta}E_{\rm hop}(t)). \label{eq:negative_epsilon_pop_deriv}
\end{equation}
Physically these two terms correspond to trajectories that have $S_z(t)$ infinitesimally greater than zero for which $S_z(t)$ is increasing (i.e.\ trajectories that have just hopped up), and trajectories with enough energy to hop that have $S_z(t)$ infinitesimally less than zero for which $S_z(t)$ is decreasing (i.e.\ trajectories which have just hopped down).

Following the same line of reasoning it can be shown that taking the limit ($\epsilon\to0^{+}$) leads instead to 
\begin{equation}
    \frac{\mathrm{d}}{\mathrm{d}t} h(S_z(t))=\delta_+(S_z(t))h(-\dot{S}_z(t))\dot{S}_z(t)
    +\delta_-(S_z(t))h(\dot{S}_z(t))\dot{S}_z(t) h(\mathit{\Delta}E_{\rm hop}(t)). \label{eq:positive_epsilon_pop_deriv}
\end{equation}
One can therefore equally weight these two expressions to obtain the symmetric expression
\begin{equation}
    \begin{aligned}
    &\frac{\mathrm{d}}{\mathrm{d}t} h(S_z(t))
    =\frac{\dot{S}_z(t)}{2}[\delta_{+}(S_z(t))+\delta_{-}(S_z(t))h\left(\mathit{\Delta}E_{\rm hop}(t)\right) ]. \label{eq:symmetric_epsilon_pop_deriv}
    \end{aligned}
\end{equation}

Each of these three expressions (Eqs.~\ref{eq:negative_epsilon_pop_deriv}, \ref{eq:positive_epsilon_pop_deriv} and \ref{eq:symmetric_epsilon_pop_deriv}) are equally valid, although they may be more or less practical. Considering first the symmetric definition, we see that inserting Eq.~\ref{eq:symmetric_epsilon_pop_deriv} into Eq.~\ref{eq:general_expectation_of_P_p_dot} and making use of the time-inversion symmetry gives
\begin{equation}
\begin{aligned}
    \frac{\mathrm{d}}{\mathrm{d}t}\langle {P}^{(a)}_p(t) \rangle =&\frac{1}{2}  \frac{\tr\left[e^{-\beta E} \delta_+(S_z)\dot{S}_z h(S_z(t)) |S_z(t)| \right]}{\tr\left[e^{-\beta E}h(S_z) |S_z|\right]}\\+&\frac{1}{2}\frac{\tr\left[e^{-\beta E}\delta_-(S_z) \dot{S}_z h\left(\mathit{\Delta}E_{\rm hop}\right) h(S_z(t)) |S_z(t)| \right]}{\tr\left[e^{-\beta E}h(S_z) |S_z|  \right]}. \label{eq:adiabatic_side_flux}
\end{aligned}
\end{equation}
This expression has the advantage that it is symmetric, although numerical implementation will involve trajectories that hop during the first time step, which one might want to avoid. Alternatively, therefore, one can make use of Eq.~\ref{eq:positive_epsilon_pop_deriv}, which has the hop infinitesimally after $t$.  This might seem like it is the wrong choice, but after inserting into Eq.~\ref{eq:general_expectation_of_P_p_dot} and making use of the time-inversion symmetry the hop will be infinitesimally before $t=0$, resulting in the following expression
\begin{equation}
\begin{aligned}
    \frac{\mathrm{d}}{\mathrm{d}t}\langle {P}^{(a)}_p(t) \rangle =&  \frac{\tr\left[e^{-\beta E} \delta_+(S_z) h(\dot{S}_z)\dot{S}_z  h(S_z(t)) |S_z(t)| \right]}{\tr\left[e^{-\beta E}h(S_z) |S_z|\right]}\\+&\frac{\tr\left[e^{-\beta E}\delta_-(S_z) h(-\dot{S}_z) \dot{S}_z h\left(\mathit{\Delta}E_{\rm hop}\right) h(S_z(t)) |S_z(t)| \right]}{\tr\left[e^{-\beta E}h(S_z) |S_z|  \right]}. \label{eq:adiabatic_side_flux_2}
\end{aligned}
\end{equation}

\subsubsection{General definition (used in main paper).}
\begin{figure}[t]
    \centering
    \includegraphics[width=0.5\linewidth]{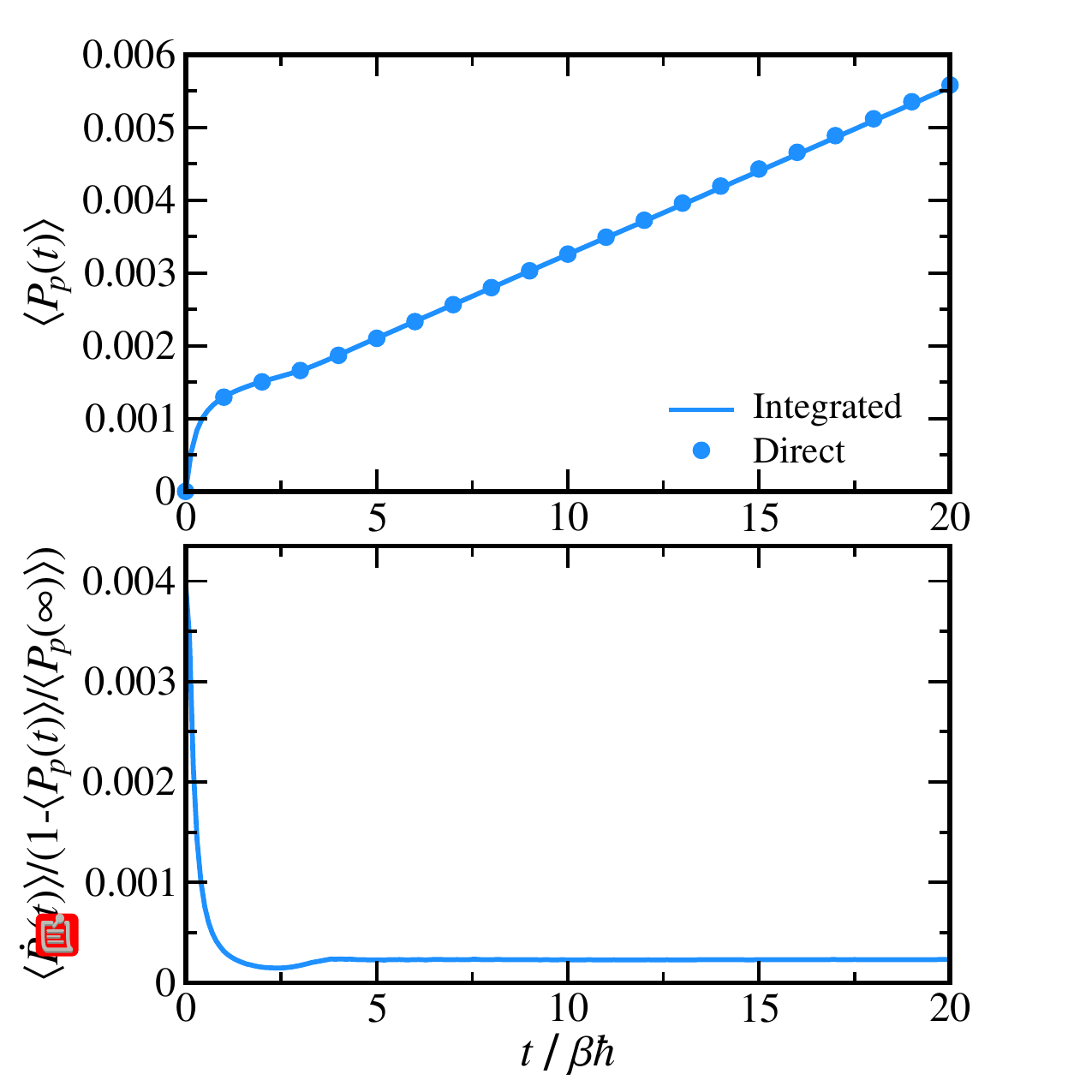}
    \caption{Demonstration of the MASH flux-correlation formalism. Upper panel: comparison of direct population dynamics (Eq.~\ref{eq:prod_MASH}) and the cumulative integral of the flux-correlation formalism (Eq.~\ref{eq:general_side_flux}) for the symmetric, $\beta\varepsilon=0$, spin-boson model with $\log_{10}(\beta\Delta)=-1$, $\beta\hbar\Omega=1/4$, $\gamma=\Omega$, and $\beta\Lambda=12$. Lower panel: generalised flux-correlation function calculated using the flux-correlation formalism (Eq.~\ref{eq:general_side_flux}) the long-time limit of which defines the rate constant. All results use the same spin-boson model, and the general definition of reactants and products.}
    \label{fig:population_vs_flux_SI}
\end{figure}

\begin{figure}[t]
    \centering
    \includegraphics[width=0.55\linewidth]{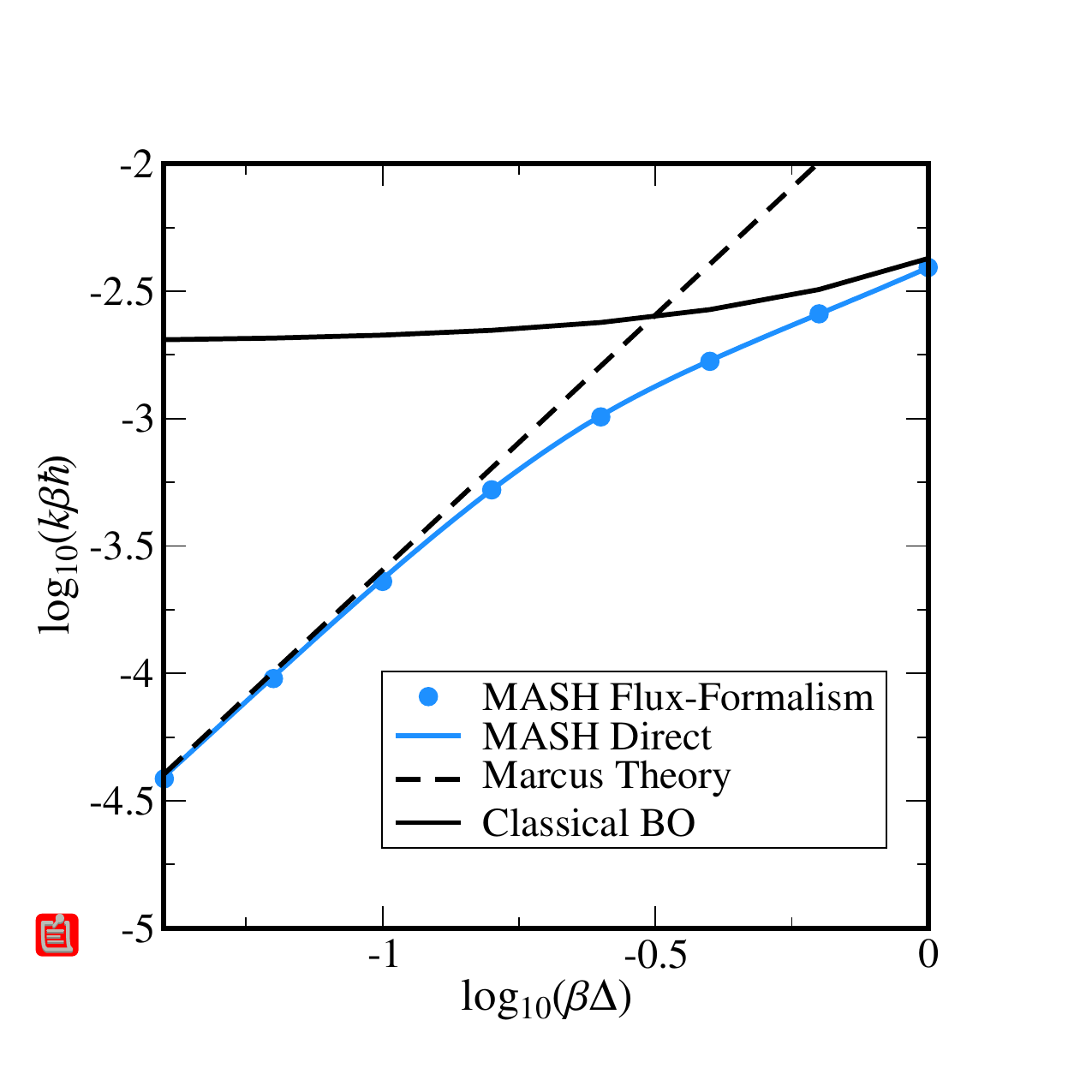}
    \caption{Log-log plot of the rate against the diabatic coupling, for  symmetric, $\beta\varepsilon=0$, spin-boson model, with $\beta\hbar\Omega=1/4$, $\gamma=\Omega$ and $\beta\Lambda=12$. Direct MASH rates were calculated from the slope of $\langle P_p(t)\rangle$, at the plateau time, between $t=10\beta\hbar$ and $t=20\beta\hbar$. Results from the flux-formulation were calculated using Eq.~\ref{eq:general_side_flux}.}
    \label{fig:symmetric_short_time_rates_delta_plot_SI}
\end{figure}

Combining the results from the previous two sections we see that the general definition of reactants and products used in the main paper (Eqs.~\ref{Product_reactant_defintions} and \ref{eq:product_defintion}) can be equivalently written as
\begin{equation}
    P_r^{(g)} = h(-(Q-Q^\ddagger))h(-S_z)+h(Q-Q^\ddagger)h(S_z)
\end{equation}
\begin{equation}
     P_p^{(g)} = h(Q-Q^\ddagger)h(-S_z)+h(-(Q-Q^\ddagger))h(S_z).
\end{equation}
 Taking the derivative with respect to time, making use of Eqs.~\ref{eq:position_derivative} and \ref{eq:symmetric_epsilon_pop_deriv}, gives
\begin{equation}
\begin{aligned}
     \dot{P}_p^{(g)} &= \dot{Q}\delta(Q-Q^\ddagger)[h(-S_z)-h(S_z)]\\
     &+\frac{1}{2}[h(-(Q-Q^\ddagger))- h(Q-Q^\ddagger)][\delta_{+}(S_z)\dot{S}_z
    +\delta_{-}(S_z(t))\dot{S}_zh(\mathit{\Delta}E_{\rm hop})].
\end{aligned}
\end{equation}
Inserting this into Eq.~\ref{eq:general_expectation_of_P_p_dot} and making use of the time-inversion symmetry we obtain
\begin{equation}
\begin{aligned}
    \frac{\mathrm{d}}{\mathrm{d}t}\langle {P}^{(g)}_p(t) \rangle =&  \frac{\tr\left[e^{-\beta E}P^{(g)}_r(t) |S_z(t)| \dot{Q}\delta(Q-Q^\ddagger)[h(S_z)-h(-S_z)]\right]}{\tr\left[e^{-\beta E}[h(-(Q-Q^\ddagger))h(-S_z)+h(Q-Q^\ddagger)h(S_z)]|S_z|\right]}\\+&  \frac{\tr\left[e^{-\beta E}P^{(g)}_r(t) |S_z(t)| [h(Q-Q^\ddagger)- h(-(Q-Q^\ddagger))]\delta_+(S_z)\dot{S}_z\right]}{2\tr\left[e^{-\beta E}[h(-(Q-Q^\ddagger))h(-S_z)+h(Q-Q^\ddagger)h(S_z)]|S_z|\right]}\\+&\frac{\tr\left[e^{-\beta E}P^{(g)}_r(t) |S_z(t)| [h(Q-Q^\ddagger)- h(-(Q-Q^\ddagger))]\delta_-(S_z)\dot{S}_z h\left(\mathit{\Delta}E_{\rm hop}\right)\right]}{2\tr\left[e^{-\beta E}[h(-(Q-Q^\ddagger))h(-S_z)+h(Q-Q^\ddagger)h(S_z)]|S_z|  \right]}.  \label{eq:general_side_flux}
\end{aligned}
\end{equation}
Here we have chosen to use the symmetric definition of the derivative of $h(S_z(t))$, however, we note that one can also obtain a similar expression that puts the hops infinitesimally before $t=0$, as was done in Eq.~\ref{eq:adiabatic_side_flux_2}.
Figures \ref{fig:population_vs_flux_SI} and \ref{fig:symmetric_short_time_rates_delta_plot_SI} show numerically that this is equivalent to the direct population dynamics used in the main text. The numerical methodology used to calculate these results is discussed in the next section.

\subsection{Numerical implementation of flux-correlation formalism.}

\subsubsection{Position based definition (normal regime).}
With some straightforward algebraic manipulations, Eq.~\ref{eq:position_side_flux} can be decomposed in  the usual Bennett--Chandler form\cite{MolSim} as follows
\begin{equation}
\begin{aligned}
    \langle \dot{P}_p(t) \rangle = & -\frac{\tr\left[e^{-\beta E}\dot{{Q}}\delta({Q}-Q^\ddagger) |{S}_z(t)| {P}_r(t)\right]}{\tr\left[e^{-\beta E}\delta({Q}-Q^\ddagger) \right]}\frac{\tr\left[e^{-\beta E}\delta({Q}-Q^\ddagger) \right]}{\tr\left[e^{-\beta E}h(-(Q-Q^\ddagger)) |S_z|\right]}\\
    =& -\left\langle \dot{Q}|{S}_z(t)| {P}_r(t)\right\rangle_{Q(0)=Q^\ddagger}\frac{2\tr\left[(e^{-\beta H_+}+e^{-\beta H_-})\delta(Q-Q^\ddagger)\right]}{\tr\left[(e^{-\beta H_+}+e^{-\beta H_-})h(-(Q-Q^\ddagger))\right]}\\
    =&-\left\langle \dot{Q}|{S}_z(t)| {P}_r(t)\right\rangle_{Q(0)=Q^\ddagger}\frac{2\left\langle\delta(Q-Q^\ddagger)\right\rangle_{\rm MF}}{\left\langle h(-(Q-Q^\ddagger))\right\rangle_{\rm MF}},
    % \langle \dot{P}_p(t) \rangle = & -\frac{\tr\left[e^{-\beta E}\dot{{Q}}\delta({Q}-Q^\ddagger) |{S}_z(t)| {P}_r(t)\right]}{\tr\left[e^{-\beta E}\delta({Q}-Q^\ddagger) \right]}\frac{\tr\left[e^{-\beta E}\delta({Q}-Q^\ddagger) \right]}{\tr\left[e^{-\beta E}h(-(Q-Q^\ddagger)) |S_z|\right]}\\
    % =& -\left\langle \dot{Q}|{S}_z(t)| {P}_r(t)\right\rangle_{Q(0)=Q^\ddagger}\frac{2\tr_{\rm nuc}\left[(e^{-\beta H_+}+e^{-\beta H_-})\delta(Q-Q^\ddagger)\right]}{r_s\tr_{\rm nuc}\left[(e^{-\beta H_+}+e^{-\beta H_-})h(-(Q-Q^\ddagger))\right]}.
    % =&-\left\langle \dot{Q}|{S}_z(t)| {P}_r(t)\right\rangle_{Q(0)=Q^\ddagger}\frac{2}{r_s}\left\langle \delta(Q-Q^\ddagger) \right\rangle_{Q<Q^\ddagger}.
\end{aligned}
\end{equation}
where 
$\langle\dots\rangle_{Q(0)=Q^\ddagger}$ indicates an average over the distribution
\begin{equation}
    \rho_{Q(0)=Q^\ddagger}(\bm{p},\bm{q},\bm{S}) = \frac{e^{-\beta E(\bm{p},\bm{q},\bm{S})}\delta({Q}-Q^\ddagger)}{\tr[e^{-\beta E(\bm{p},\bm{q},\bm{S})}\delta({Q}-Q^\ddagger)]},
\end{equation}
and $\langle\dots\rangle_{\rm MF}$ indicates an average over the thermal ``Mean-Field'' distribution 
\begin{equation}
\rho_{\rm MF}(\bm{p},\bm{q}) = \frac{e^{-\beta H_+}+e^{-\beta H_-}}{\tr[e^{-\beta H_+}+e^{-\beta H_-}]}.
\end{equation}
Note that in this form, it is assumed that $S_z$ is sampled with the correct Boltzmann weights. Alternatively, we can sample $\bm{S}$ uniformly from the Bloch sphere and write  
\begin{equation}
\begin{aligned}
    \langle \dot{P}_p(t) \rangle 
    =& -\left\langle \frac{2e^{-\beta E}}{e^{-\beta H_+}+e^{-\beta H_-}}\dot{Q}|{S}_z(t)| {P}_r(t)\right\rangle_{Q(0)=Q^\ddagger,{\rm MF
}}\frac{2\left\langle\delta(Q-Q^\ddagger)\right\rangle_{\rm MF}}{\left\langle h(-(Q-Q^\ddagger))\right\rangle_{\rm MF}},
    %     \langle \dot{P}_p(t) \rangle 
    % =& -\left\langle \frac{2e^{-\beta E}}{e^{-\beta H_+}+e^{-\beta H_-}}\dot{Q}|{S}_z(t)| {P}_r(t)\right\rangle_{Q(0)=Q^\ddagger,\bm{S}_{\rm uniform}}\\&\times\frac{2\tr_{\rm nuc}\left[(e^{-\beta H_+}+e^{-\beta H_-})\delta(Q-Q^\ddagger)\right]}{r_s\tr_{\rm nuc}\left[(e^{-\beta H_+}+e^{-\beta H_-})h(-(Q-Q^\ddagger))\right]}.
    % =&-\left\langle \dot{Q}|{S}_z(t)| {P}_r(t)\right\rangle_{Q(0)=Q^\ddagger}\frac{2}{r_s}\left\langle \delta(Q-Q^\ddagger) \right\rangle_{Q<Q^\ddagger}.
\end{aligned}
\end{equation}
where $\langle\dots\rangle_{Q(0)=Q^\ddagger,{\rm MF}}$ indicates an average over the distribution 
\begin{equation}
\rho_{Q(0)=Q^\ddagger,{\rm MF}}(\bm{p},\bm{q}) = \frac{(e^{-\beta H_+}+e^{-\beta H_-})\delta({Q}-Q^\ddagger)}{\tr[(e^{-\beta H_+}+e^{-\beta H_-})\delta({Q}-Q^\ddagger)]}.
\end{equation}

\subsubsection{Adiabatic state defintion (inverted regime).}
We can evaluate Eq.~\ref{eq:adiabatic_side_flux} using a modified Bennett--Chandler procedure. We start by noting that the presence of the derivative coupling in
\begin{equation}
    \dot{S}_{z}=-\sum_{j}\frac{2d_{j}(\bm{q})p_{j}}{m_j}S_{x}
\end{equation}
means that trajectories starting in regions of high derivative coupling will dominate. Ideally, one should therefore incorporate this into the sampling. Practically for our purposes, the sampling of the nuclear positions and momenta can be done from an as yet undefined distribution
\begin{equation}
    \rho_{\rm samp}(\bm{p},\bm{q}) = \frac{e^{-\beta H_{\rm samp}(\bm{p},\bm{q}) }}{\tr\left[e^{-\beta H_{\rm samp}(\bm{p},\bm{q}) }\right]}.
\end{equation}
The first term of Eq.~\ref{eq:adiabatic_side_flux} can be written in terms of averages over this distribution as
\begin{multline}
     \frac{\tr\left[e^{-\beta E} \delta_+(S_z)\dot{S}_z h(S_z(t)) |S_z(t)| \right]}{\tr\left[e^{-\beta E}h(S_z) |S_z|\right]}
     \\=\left\langle e^{-\beta (E-H_{\rm samp})} \dot{S}_z h(S_z(t)) |S_z(t)|\right\rangle_{{\rm samp},S_z=0^+} \frac{\tr\left[e^{-\beta H_{\rm samp}}\delta_+(S_z)\right]}{\tr\left[e^{-\beta E}h(S_z) |S_z|\right]}. 
\end{multline}
The integrals over the spin-sphere in the second term can be performed analytically to give
\begin{equation}
\begin{aligned}
    \frac{\tr\left[e^{-\beta H_{\rm samp}}\delta_+(S_z)\right]}{\tr\left[e^{-\beta E}h(S_z) |S_z|\right]} 
    &= \frac{\tr\left[e^{-\beta H_{\rm samp}}\right]\int_{-1}^1 \mathrm{d}S_z\,\delta_+(S_z)}{\tr\left[e^{-\beta H_+}\right]\int_{-1}^1 \mathrm{d}S_z \,h(S_z)| S_z|} \\
    % &=\frac{1}{\langle e^{-\beta(H_+-H_{\rm samp})}\rangle_{\rm samp}} \frac{\int_{-1}^1 \mathrm{d}u\,\delta(u)}{\int_{-1}^1 \mathrm{d}u \,h(u)| u|}\\
    &=\frac{2}{\langle e^{-\beta(H_+-H_{\rm samp})}\rangle_{\rm samp}} ,
\end{aligned}    
\end{equation}
such that overall we have
\begin{equation}
    \begin{aligned}
     &\frac{\tr\left[e^{-\beta E} \delta_+(S_z)\dot{S}_z h(S_z(t)) |S_z(t)| \right]}{\tr\left[e^{-\beta E}h(S_z) |S_z|\right]}= \frac{2 \left\langle e^{-\beta (E-H_{\rm samp})} \dot{S}_z h(S_z(t)) |S_z(t)|\right\rangle_{{\rm samp},S_z=0^+}}{\langle e^{-\beta(H_+-H_{\rm samp})}\rangle_{\rm samp}} . 
\end{aligned}
\end{equation}
For the term constrained initially to the lower adiabatic surface, we can follow the same steps to obtain
the final equation for the derivative of the population as
\begin{equation}
\begin{aligned}
    \frac{\mathrm{d}}{\mathrm{d}t}\langle {P}^{(a)}_p(t) \rangle =&   \frac{\left\langle e^{-\beta (E-H_{\rm samp})} \dot{S}_z h(S_z(t)) |S_z(t)|\right\rangle_{{\rm samp},S_z=0^+}}{\langle e^{-\beta(H_+-H_{\rm samp})}\rangle_{\rm samp}} \\+& \frac{\left\langle e^{-\beta (E-H_{\rm samp})} \dot{S}_zh\left(\mathit{\Delta}E_{\rm hop}\right) h(S_z(t)) |S_z(t)|\right\rangle_{{\rm samp},S_z=0^-}}{\langle e^{-\beta(H_+-H_{\rm samp})}\rangle_{\rm samp}} . 
\end{aligned}
\end{equation}
Here we give the expression for the symmetric definition of the derivative, $\dot{P}_p^{(a)}$, however a similar expression can also be derived for Eq.~\ref{eq:adiabatic_side_flux_2}. Note that sampling from the distribution with $S_z(0)=0^+$ or $S_z(0)=0^-$ is implemented in practice by just setting the initial $S_z$ to a small floating point number, $\pm10^{-10}$.

\subsubsection{General definition (used in main paper).}

Combining the ideas of the last two sections, we can arrive at a Bennett--Chandler scheme to efficiently compute the rate with the general dividing surface. For the first term in Eq.~\ref{eq:general_side_flux}, this can be achieved by writing
\begin{equation}
\begin{aligned}
     &\frac{\tr\left[e^{-\beta E}\dot{Q}\delta(Q-Q^\ddagger)[h(S_z)-h(-S_z)]P^{(g)}_r(t) |S_z(t)| \right]}{\tr\left[e^{-\beta E}[h(-(Q-Q^\ddagger))h(-S_z)+h(Q-Q^\ddagger)h(S_z)]|S_z|\right]}
\\&=\left\langle \dot{Q}[h(S_z)-h(-S_z)]P^{(g)}_r(t) |S_z(t)| \right\rangle_{Q(0)=Q^\ddagger} \frac{2\tr\left[(e^{-\beta H_{+}}+ e^{-\beta H_-})\delta(Q-Q^\ddagger)\right]}{ \tr\left[e^{-\beta H_{+}}h(Q-Q^\ddagger)+ e^{-\beta H_-}h(-(Q-Q^\ddagger))\right]}
\\&= \frac{2\left\langle \dot{Q}[h(S_z)-h(-S_z)]P^{(g)}_r(t) |S_z(t)| \right\rangle_{Q(0)=Q^\ddagger}\langle \delta(Q-Q^\ddagger \rangle_{\rm MF}}{ \left\langle \frac{e^{-\beta H_{+}}}{e^{-\beta H_{+}}+ e^{-\beta H_-}}h(Q-Q^\ddagger)+ \frac{e^{-\beta H_-}}{e^{-\beta H_{+}}+ e^{-\beta H_-}}h(-(Q-Q^\ddagger))\right\rangle_{\rm MF}}. 
\end{aligned}
\end{equation}
Alternatively sampling $\bm{S}$ uniformly from the surface of a sphere, we have that
\begin{multline}
    \left\langle \dot{Q}[h(S_z)-h(-S_z)]P^{(g)}_r(t) |S_z(t)| \right\rangle_{Q(0)=Q^\ddagger} \\=\left\langle \frac{2e^{-\beta E}}{e^{-\beta H_+}+e^{-\beta H_-}} \dot{Q}[h(S_z)-h(-S_z)]P^{(g)}_r(t) |S_z(t)| \right\rangle_{Q(0)=Q^\ddagger,{\rm MF}}. 
\end{multline}

For the second term and third term, we can follow the same approach as with the pure adiabatic dividing surface. This gives 
\begin{multline}
      \frac{\tr\left[e^{-\beta E} [h(Q-Q^\ddagger)- h(-(Q-Q^\ddagger))]\delta_+(S_z)\dot{S}_zP^{(g)}_r(t) |S_z(t)|\right]}{2\tr\left[e^{-\beta E}[h(-(Q-Q^\ddagger))h(-S_z)+h(Q-Q^\ddagger)h(S_z)]|S_z|\right]}
\\
      = \frac{\left\langle e^{-\beta (E-H_{\rm samp})} [h(Q-Q^\ddagger)- h(-(Q-Q^\ddagger))] \dot{S}_z P^{(g)}_r(t) |S_z(t)|\right\rangle_{{\rm samp},S_z=0^+}}{ \left\langle e^{-\beta (H_--H_{\rm samp})}h(-(Q-Q^\ddagger))   +e^{-\beta (H_+-H_{\rm samp})}h(Q-Q^\ddagger) \right\rangle_{{\rm samp}}}
\end{multline}
and
\begin{multline}
    \frac{\tr\left[e^{-\beta E} [h(Q-Q^\ddagger)- h(-(Q-Q^\ddagger))]\delta_-(S_z)\dot{S}_z h\left(\mathit{\Delta}E_{\rm hop}\right) P^{(g)}_r(t) |S_z(t)|\right]}{2\tr\left[e^{-\beta E}[h(-(Q-Q^\ddagger))h(-S_z)+h(Q-Q^\ddagger)h(S_z)]|S_z|  \right]}
\\= \frac{\left\langle e^{-\beta (E-H_{\rm samp})} [h(Q-Q^\ddagger)- h(-(Q-Q^\ddagger))] h(\mathit{\Delta}E_{\rm hop})\dot{S}_z P^{(g)}_r(t) |S_z(t)| \right\rangle_{{\rm samp},S_z=0^-}}{ \left\langle e^{-\beta (H_--H_{\rm samp})}h(-(Q-Q^\ddagger))   +e^{-\beta (H_+-H_{\rm samp})}h(Q-Q^\ddagger) \right\rangle_{{\rm samp}}}. 
\end{multline}
Here we have chosen to use the symmetric definition of $\frac{\mathrm{d}}{\mathrm{d}t}h(S_z(t))$ (Eq.~\ref{eq:symmetric_epsilon_pop_deriv}). This means that hops can occur within the first time step, however we found that this did not present a significant numerical difficulty. We found that it was more important to sample the initial momenta such that for each $\bm{p}$, we also sampled a trajectory with $-\bm{p}$ at the same $\bm{S}$ and $\bm{q}$. We note that this could have also been achieved by using the asymmetric version, where hops are always infinitesimally before $t=0$, by appropriately rescaling the momenta. Finally in the present calculations, we used a very simple choice for $H_{\rm samp}$ corresponding to sampling from the reactant diabatic potential origin, shifted to the diabatic crossing seam
\begin{equation}
    H_{\rm samp}(\bm{p},\bm{q}) = \sum_{j=1}^f \frac{p_j^2}{2m_j} + U_0(\bm{q}-\bm{q}_{\rm shift})
\end{equation}
with $\bm{q}_{\rm shift}=(Q^\ddagger,0,0,0,\dots)$. This is not the optimal choice, as it does not take into account the narrowing of the derivative coupling as $\Delta\to0$, however it was sufficient for the present purpose.

\section{Additional Results.}
In the following, we include results of additional systems, that support the conclusions of the main text.

\begin{figure}[t]
    \centering
    \includegraphics[width=0.5\linewidth]{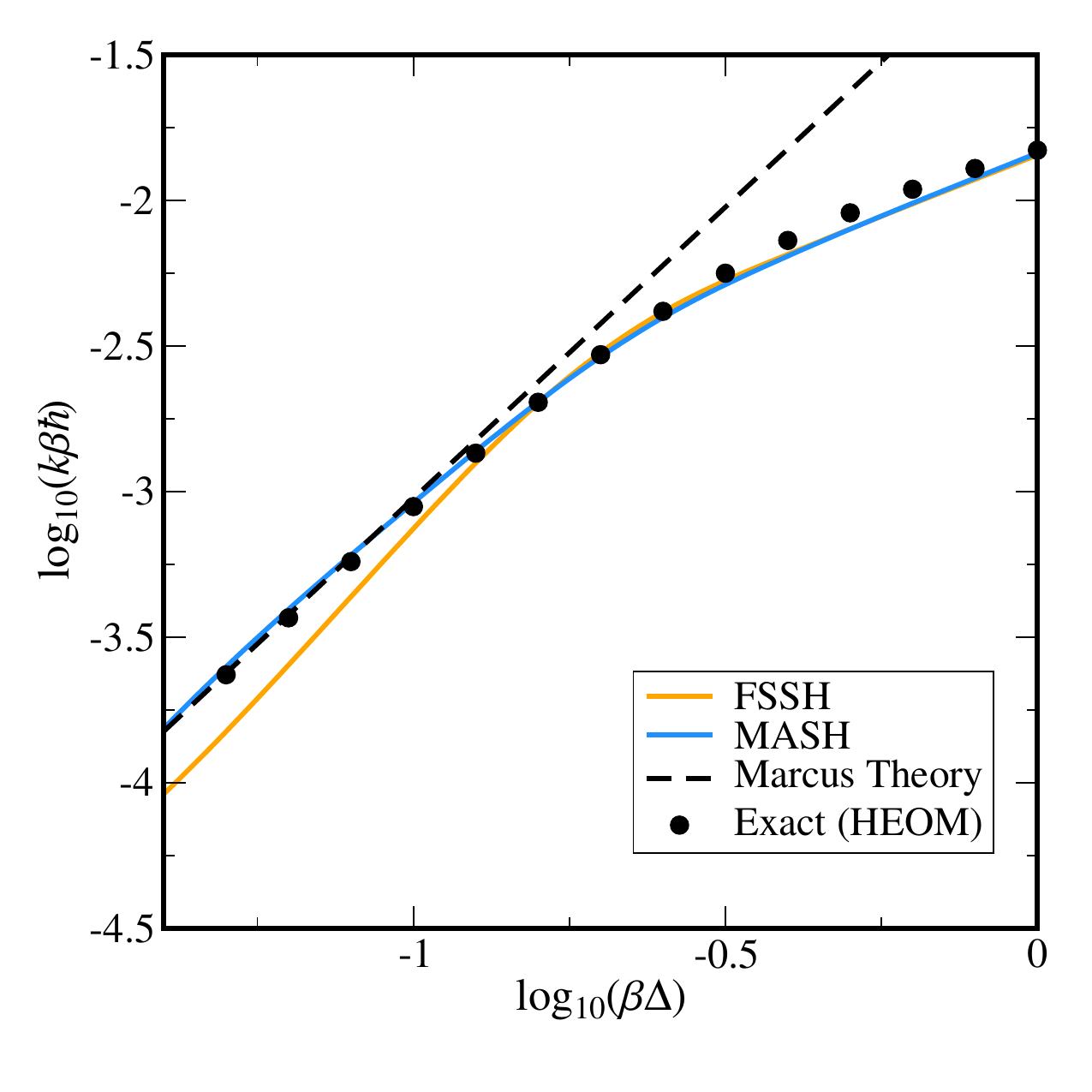}
    \caption{Log-log plot of the rate against the diabatic coupling, for an asymmetric, $\beta\varepsilon=3$, spin-boson model, with $\beta\hbar\Omega=1/4$, $\gamma=\Omega$ and $\beta\Lambda=12$. FSSH and MASH rates were calculated from the slope of $\langle P_p(t)\rangle$, at the plateau time, between $t=10\beta\hbar$ and $t=20\beta\hbar$. This shows essentially the same behaviour as seen  for the symmetric system in Fig.~1 
 of the main text.}
    \label{fig:asymmetric_short_time_rates_delta_plot_SI}
\end{figure}

In Fig.~\ref{fig:asymmetric_short_time_rates_delta_plot_SI}, we show that the qualitative behaviour is unchanged from Fig.~1 
 of the main text when one adds a small bias to products, $\beta\varepsilon=3$.

\begin{figure}[t]
    \centering
    \includegraphics[width=0.5\linewidth]{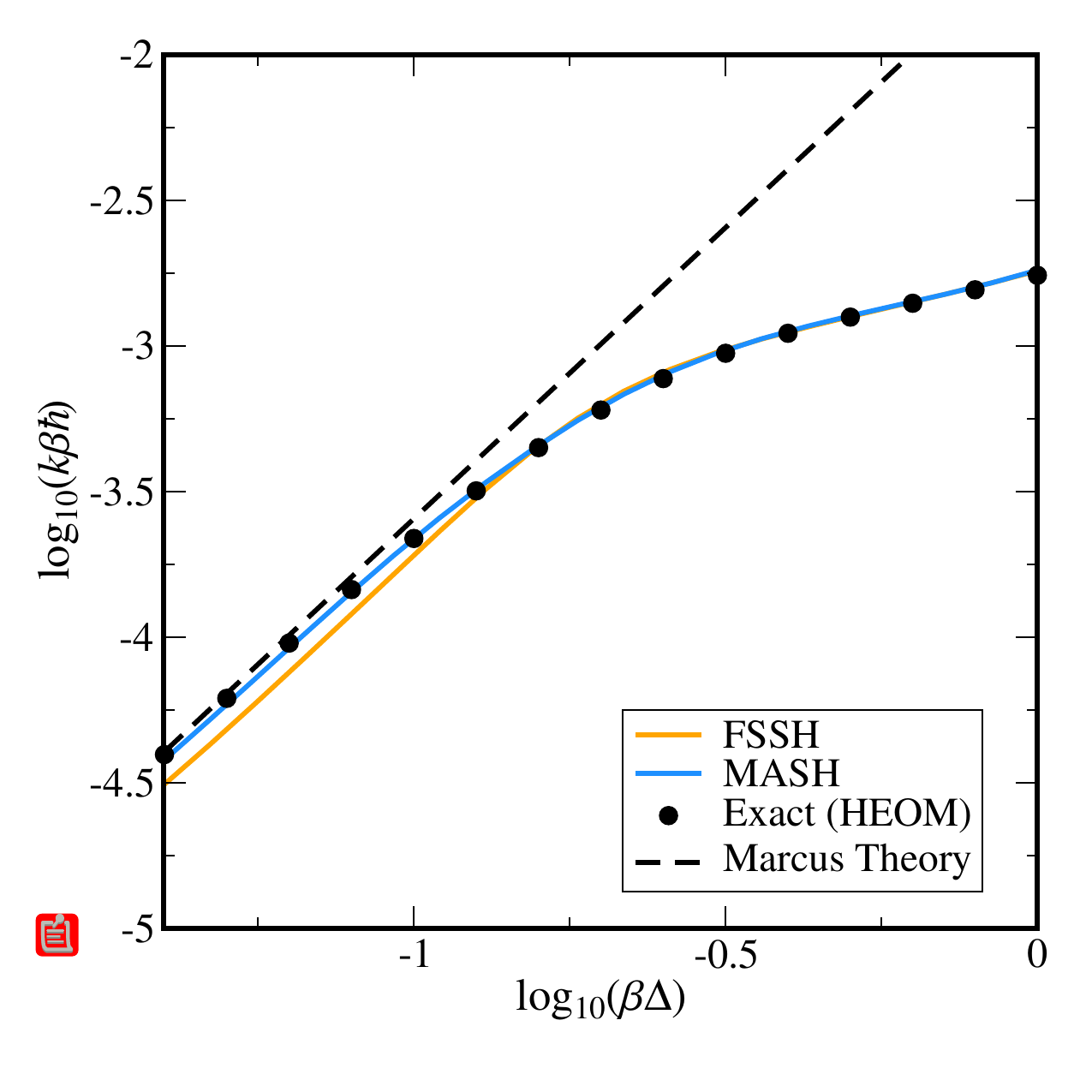}
    \caption{Log-log plot of the rate against the diabatic coupling, for a symmetric, $\beta\varepsilon=0$, spin-boson model, with $\beta\hbar\Omega=1/4$, and $\beta\Lambda=12$ in the overdamped regime $\gamma=4\Omega$. FSSH and MASH rates were calculated from the slope of $\langle P_p(t)\rangle$, at the plateau time, between $t=10\beta\hbar$ and $t=20\beta\hbar$.}
    \label{fig:symmetric_short_time_rates_overdamped_delta_plot_SI}
\end{figure}

\begin{figure}[t]
    \centering
    \includegraphics[width=0.5\linewidth]{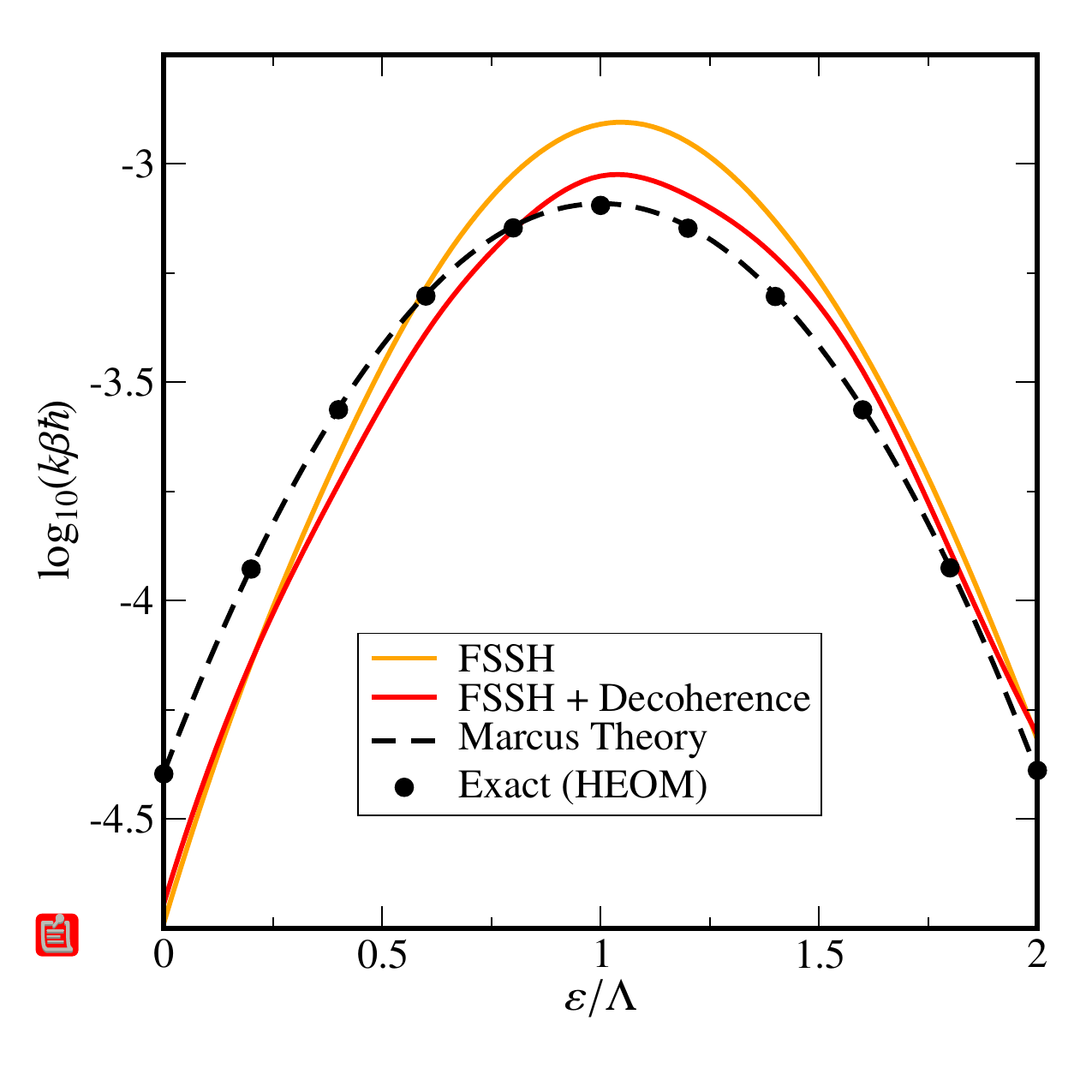}
    \caption{Logarithmic plot of the rate against reaction driving force, showing the famous Marcus turn over behaviour for a spin-boson model with weak diabatic coupling, $\log_{10}(\beta\Delta)=-7/5$, $\beta\hbar\Omega=1/4$, $\gamma=\Omega$, and $\beta\Lambda=12$. FSSH rates were calculated from the slope of $\langle P_p(t)\rangle$, at the plateau time, between $t=10\beta\hbar$ and $t=20\beta\hbar$.
    To obtain the red line, the simple gap-based decoherence correction was applied whenever $V_+-V_->4k_{\rm B}T$.}
    \label{fig:inverted_regime_plot_SI}
\end{figure}

\begin{figure}[t]
    \centering
    \includegraphics[width=0.5\linewidth]{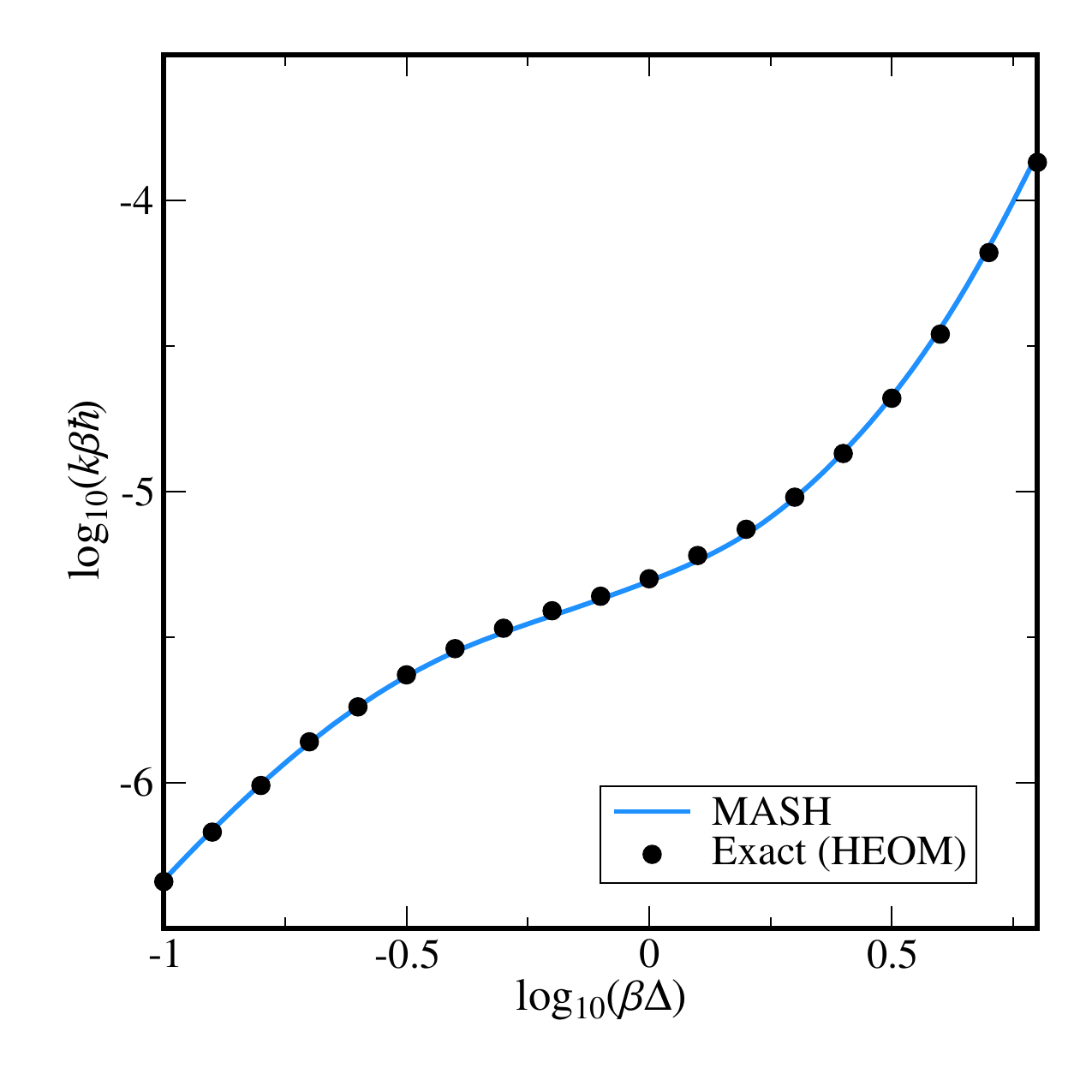}
    \caption{Log-log plot of the rate against the diabatic coupling, for an asymmetric, $\varepsilon=\Lambda/4$, spin-boson model, with $\beta\hbar\Omega=1/2$, $\gamma=32\Omega$ and $\beta\Lambda=60$. MASH rates were calculated using the flux-correlation formalism, and exact HEOM results were taken from Ref.~\citenum{Lawrence2019ET}.}
    \label{fig:asymmetric_large_reorg_rates_delta_plot_SI}
\end{figure}

In Fig.~\ref{fig:symmetric_short_time_rates_overdamped_delta_plot_SI}, we consider a system which is equivalent to that considered in Fig.~1 
of the main text, except that it is in the overdamped $\gamma=4\Omega$ rather than underdamped regime. This illustrates that the overcoherence problem is less significant at high friction.

In Fig.~\ref{fig:inverted_regime_plot_SI}, we show the results of including the simple decoherence correction on FSSH. These were left out of the main paper to avoid clutter. This illustrates that unlike for MASH, the simple decoherence correction is unable to fix FSSH.

In Fig.~\ref{fig:asymmetric_large_reorg_rates_delta_plot_SI}, we consider a system with a large reorganisation energy, $\beta\Lambda=60$, and high friction, $\gamma=32\Omega$. This illustrates the utility of the flux-correlation formalism, as direct calculation of the rates would be prohibitively expensive at the smallest values of $\Delta$ considered.

\clearpage
\bibliographystyle{achemso}
\bibliography{references,extra_refs} %,references}